\documentstyle[preprint,aps]{revtex}

\begin{document}

\title{{\large\bf  Anderson-Hubbard Model in $d = \infty$}}

\author{{\bf  M. Ulmke$^{a}$,  V. Jani\v{s}$^{b,}$\footnote[1]{Permanent
address:
Institute of Physics, Academy of Sciences of the Czech Republic, \\
\hspace*{16pt} CZ-18040 Praha 8, Czech Republic},
 and D.~Vollhardt$^{b}$}\\
\vspace{10pt}
{\it $^{(a)}$ Institut f\"{u}r Festk\"{o}rperforschung, Forschungszentrum
J\"{u}lich,
52425 J\"{u}lich, \\
Federal Republic of Germany\\
$^{(b)}$Institut f\"{u}r Theoretische Physik C,
Technische Hochschule Aachen,
52056 Aachen, Federal Republic of Germany}}

\date{\today}
\maketitle
\newpage

\begin{abstract}
We present a detailed, quantitative study of the competition between
interaction- and disorder-induced effects in
electronic systems. For this the
Anderson-Hubbard model with diagonal disorder is investigated analytically
and numerically in the limit of infinite
spatial dimensions, i.e.~within a dynamical
mean-field theory, at half filling. Numerical results are obtained  for three
different disorder distributions by employing Quantum Monte Carlo techniques
which provide an explicit finite-temperature solution of the model
in this limit. The magnetic phase diagram is constructed from the
zeros of the inverse
averaged staggered susceptibility. We find that at low enough temperatures and
sufficiently strong interaction there always exists
 a phase with antiferromagnetic
long-range order. A novel strong coupling anomaly,
 i.e.~an {\it increase} of the N\'{e}el-temperature for increasing disorder,
is discovered. An explicit explanation is given which shows that in the
case of diagonal disorder this is a generic effect. -- The existence of
metal-insulator
transitions is studied by evaluating the averaged compressibility both in the
paramagnetic and antiferromagnetic phase. A rich transition
scenario, involving
metal-insulator and magnetic transitions, is found and its dependence
on the choice of the disorder distribution is discussed.
\end{abstract}

\newpage

\section{Introduction}

The investigation of interacting electronic systems is one
of the most intriguing, albeit difficult, subjects in condensed
matter physics. The same is true for the study of disordered,
non-interacting electrons.
In view of the theoretical complexity of the two problems taken separately
it is understandable that their combination, i.e.~the
{\em simultaneous} presence of randomness  and interactions, as found
in many real systems (e.g. doped semiconductors near the metal-insulator
transition, high-$T_c$ superconducting materials close to $T_c$, etc.),
leads to new, fundamental questions to which only few secured answers are
known. This is all the more true when the interaction and/or the disorder
is strong, since there exist hardly any tractable,
and at the same time controlled,
 theoretical method of investigation in this
limit.

An important starting point for the investigation
of interacting, disordered systems
was the field-theoretic approach developed for
the treatment of non-interacting
electrons, i.e.~the scaling theory of Anderson localization
\cite{1,2,3}. A generalization of this
theory to finite interactions by Finkelshtein \cite{4} provided
essential new insight \cite{5}. However,
the appearance of local magnetic moments
in the renormalization group treatment discovered by him \cite{6} and
Castellani et al. \cite{7} turned out to be a
fundamental obstacle for the study of
the metal-insulator transition (MIT) itself
\cite{8,9}.
A microscopic origin of this instability towards the
formation of localized moments can already be
traced within Hartree Fock theory for
the disordered Hubbard model (``Anderson-Hubbard model'') with off-diagonal
disorder  \cite{10,11}.
The results indicate that the above renormalization-group approach,
as well as is starting point, i.~e. Anderson-localization,
are not appropriate when it
comes to the investigation of three-dimensional disordered electrons with
intermediate or strong interactions close to half-filling ($n \simeq 1$).
After all, a MIT occurs even without disorder in this case, and the lattice
periodicity becomes essential. A non-perturbative and qualitatively
quite different, but essentially uncontrolled, renormalization technique
by which disorder and interactions can be treated on equal footing
 is the real space
renormalization group approach.
It was first applied by Ma \cite{12} and recently used by Yi et al. \cite{13}
to investigate the Anderson-Hubbard model with diagonal disorder,
i.e. random on-site
energies, in dimensions $d = 1,3$
at $n = 1$ and in $d = 2$ for $n \stackrel{<}
{\sim} 1$, respectively. However, in these investigations the formation
 of antiferromagnetic long range order (AFLRO) which will always
set in at large enough repulsion (at least at
$n = 1$) was not considered at all.
The strong coupling limit of this model was
studied by Zimanyi and Abrahams \cite{14} using a slave-boson formulation
of the corresponding $t-J$ model. The considerably simpler, but
still highly non-trivial,  case
of disordered, {\em spinless} fermions in $d = 1$ was also addressed
recently \cite{15,16}.

To obtain a global picture of the
properties of interacting, disordered
systems it is desirable to know the solution of a simple,
microscopic model which is valid for all
input parameters (interaction, disorder,
temperature, band filling).
Since exact solutions are not available in $d = 2,3$ one would
like to construct, at least, a thermodynamically
consistent mean-field theory that is
valid also at strong coupling.
Such a (non-perturbative) approximation is
provided by the exact solution of a
model in $d = \infty$.
It is now known
that even in the limit $d \to \infty$
\cite{17,18}
 the Hubbard interaction remains dynamical \cite{19}
 and leads to a highly non-trivial single-site problem
\cite{20,27}
with infinitely many coupled quantum degrees of freedom.
This problem is, in fact, equivalent with an Anderson impurity model
complemented by a self-consistency condition
\cite{21,24,27}
 and is thus amenable to numerical investigations
 within a finite-temperature quantum Monte-Carlo
approach \cite{28}. In the absence of disorder this technique
was already used
by several groups to investigate the  magnetic phase diagram
\cite{24,25,29},
 the Mott-Hubbard transition
\cite{30,32,29},
 transport properties \cite{33},
and lately also superconductivity in a two-band version \cite{34}
 of the Hubbard model in
$d = \infty$. These investigations were also extended to
 the periodic Anderson model \cite{35} and the Holstein model \cite{36}.

The dynamical mean-field theory of interacting electrons  obtained
in the limit $d = \infty$ has many appealing
 features; however, it also has its
limitations which are due to the purely local nature of the theory.
 One effect which cannot
be described within the $d = \infty$ mean-field theory of
disordered electrons is the phenomenon of Anderson localization.
That is, the electrons in infinite dimensions are delocalized
for arbitrary strength of the disorder. The absence of
localized states is caused by the above mentioned single-site dependence,
i.e.~by the
fact that momentum dependence enters the physical quantities
only through the dispersion relation
whereby vertex corrections
to the conductivity vanish identically \cite{37}.
In fact, it was  shown \cite{38,39}
that for vanishing interaction, when the Anderson-Hubbard model reduces to
the Anderson disorder model, the well-known and
much-used ``Coherent Potential Approximation (CPA)''
\cite{40,41}, which does not describe Anderson
localization, yields the {\em exact}
solution in $d = \infty$. Nevertheless
it should be stressed that even in the absence of Anderson
localization the presence of
disorder can have a significant, non-trivial effect on
the properties of an interacting system. First results of a
quantum Monte-Carlo study of the competition
between disorder and interaction in
the Anderson-Hubbard  in $d = \infty$ with diagonal ``binary alloy''
disorder were recently reported by us \cite{42}. The
importance of separable,
off-diagonal disorder in
 the formation of local moments and the Mott transition
was investigated by Dobrosavljevi\'{c} and Kotliar
\cite{43}.

A good impression of how complicated the phase diagram and the dynamical
properties of interacting electrons with and without disorder are may be
inferred from the results for spinless electron in $d = \infty$
\cite{44,45}. The fact that the interactions reduce to
their Hartree-contribution in this case \cite{19}, i.e.~are no longer
dynamical, does not at all imply that the
properties of the model become trivial -- quite
the contrary. Many details of this model can be obtained analytically in
$d = \infty$, thereby providing valuable insight into the
effects of interactions \cite{45}, the interplay between interaction and
disorder \cite{44}, and the $d = \infty$ limit itself \cite{46}.

The correlations caused by interactions
 between electrons and by the scattering of electrons in a disordered
medium, respectively, can be very different; indeed, they may
lead to  opposite effects. For example, even
an arbitrarily weak repulsive Hubbard interaction
between electrons on a bipartite
 lattice will induce AFLRO when the band filling is sufficiently
close to $n = 1$ (at least in dimensions $d > 2$).
 By contrast, it is the very nature of a random potential
to  oppose order. Hence interactions and disorder are expected to
compete with each other.
This conclusion may appear self-evident
and almost trivial
but is really  only based
on a simple-minded superposition of the two individual physical
effects (which may, in fact,
be quite correct for weak interactions and
disorder). When the interaction and/or disorder is strong, however, this
picture looses its basis. In general the mutual interplay of those two
effects can be expected to lead to
{\em novel}, non-perturbative, quantum-mechanical
many-body phenomena, e.g. new phases,
which have no analog in non-random, interacting
or disordered, non-interacting systems.

In view of the interplay between disorder and interaction
there are a number of simple questions
whose answers will be non-trivial, for
example
\begin{enumerate}
\item
Is there a {\em finite} disorder strength above which
AFLRO ceases to exist? (After all, the fluctuations
of the ensembles around
the average value $n = 1$ induced by the disorder might become so large
that AFLRO is made impossible).
\item
Does disorder always {\em reduce} the critical
temperature for AFLRO, as one might
expect intuitively?
\item
Do interactions always drive a system {\em away} from  metallic
behavior into an insulating state?
\item
How important is the choice of disorder distribution on the results?
\end{enumerate}
These are questions which need to be answered by means
of controlled investigations of a well-defined microscopic model.
It is our intention
in this paper to provide explicit, numerically exact answers to the above
questions within the dynamical mean-field theory obtained
in the limit of high spatial dimensions.
A preliminary account was presented in Ref. \cite{42}.
The paper is structured as follows. In Sect.~2 we
introduce the model and discuss the form of the corresponding
averaged free
energy in $d = \infty$. In Sect.~3
we calculate correlation functions, i.e.~the averaged compressibility
and staggered susceptibility, to determine the
 thermodynamic stability of various
phases. The details of the model (input parameters, disorder
distribution etc.) are specified  and the numerical procedure
is discussed in Sect.~4. This leads to the
construction of the magnetic phase diagram in Sect.~5.
In Sect.~6 the results on magnetic
phase transitions are then complemented by
those on metal-insulator transitions.
 A discussion of the combined phase transition scenario closes the
presentation.

\section{Model and Averaged Free Energy in $d = \infty$}

The simplest microscopic Hamiltonian for conduction electrons
 interacting via the local Hubbard interaction
in a disordered system is the so-called Anderson-Hubbard model,
which can be written as
\begin{equation}
\hat{H} =  \sum_{i,j , \sigma} t_{ij } \hat{c}_{i \sigma}^{+}
\hat{c}_{j \sigma}^{} + U \sum_i \hat{n}_{i \uparrow}
\hat{n}_{i \downarrow} + \sum_{i, \sigma}
(\epsilon_i - \mu_\sigma) \hat{n}_{i \sigma} .
\label{Gl1}
\end{equation}
Here $\hat{c}_{i \sigma}^{+} (\hat{c}_{i \sigma}^{})$ create (annihilate) a
$\sigma$-electron
on site $i$, with $\hat{n}_{i \sigma} = \hat{c}_{i \sigma}^{+}
\hat{c}_{i \sigma}^{}$
and $\mu_\sigma$ is the chemical potential of the $\sigma$-electrons.
In general
both the hopping amplitude $t_{ij }$ and the atomic energies $\epsilon_i$
in (\ref{Gl1}) depend on the configuration of the atoms distributed on the
lattice. In this paper we restrict ourselves to the case
of  uncorrelated, diagonal disorder.
Hence  we assume the hopping elements to be
independent of the randomness (with $t_{ij } = -t$ for nearest neighbor sites
and $t_{ij } = 0$ otherwise) while the atomic
potentials $\epsilon_i$ are chosen as  random, their distribution being
purely local; thus short-range order is neglected.
The Hamiltonian under investigation is then given by
\begin{equation}
\hat{H} = \hat{H}_{Hub} + \sum_{i , \sigma} (\epsilon_i - \mu_\sigma)
\hat{n}_{i \sigma}
\label{Gl2}
\end{equation}
where $\hat{H}_{Hub} $ is the usual Hubbard model
\begin{equation}
\hat{H}_{Hub} = - \frac{t^*}{\sqrt{Z}} \sum_{\langle i j \rangle, \sigma}
\hat{c}_{i \sigma}^{+}
\hat{c}_{j \sigma}^{} + U \sum_i \hat{n}_{i \uparrow} \hat{n}_{i \downarrow} .
\label{Gl3}
\end{equation}
Here the hopping matrix element was written as \cite{17}
\begin{equation}
t = t^*/\sqrt{Z}\; ,\;  t^* = const,
\label{Gl4}
\end{equation}
 with $Z$ as the
number of nearest-neighbors on the lattice (e.g.
 $Z = 2d$ on hypercubic lattices in $d$ dimensions).
This scaling ensures that
(\ref{Gl3}) remains non-trivial even in the limit $ Z \to \infty$.
We can then write the averaged full energy (grand potential) as
\begin{mathletters}
\label{Gl5}
\begin{equation}
\Omega_{av}  = - \beta^{-1} \langle
\big( \ln \;  {\rm Tr} \; \exp (- \beta \hat{H} ) \big) \rangle_{av}
\label{Gl5a}
\end{equation}
where $\beta = 1/k_BT$. Here the average over the disorder
is defined as usual as
\begin{equation}
\langle X \rangle_{av}
=   \prod_{i } \Big[ \int_{- \infty}^{\infty} d \epsilon_i
P(\epsilon_i) \Big] X (\epsilon_1 , \ldots , \epsilon_L)
\label{Gl5b}
\end{equation}
\end{mathletters}
where $L$ is the  number of lattice sites, and
$P(\epsilon_i)$ is the distribution of the random potentials
$\epsilon_i$. The
explicit form of the function $P(\epsilon_i)$
will be specified later.

In realistic dimensions $(d = 2,3)$
it is impossible to perform the trace over all fermionic
states in (\ref{Gl2}) exactly. However, in the
limit of infinite coordination number or
spatial dimensions
significant simplifications occur
in the evaluation of thermodynamic properties.
Two of us recently showed \cite{39} that the averaged grand potential
of the disordered Hubbard model in $d = \infty$ can be reduced
to an expression where the averaging is performed only on a single site. The
explicit expression for the grand potential in
the paramagnetic phase is given by
\begin{eqnarray}
\Omega_{av}/L &=& - \beta^{-1} \sum_{\sigma n} \int_{- \infty}^{\infty}
dE N (E)
\ln [ i \omega_n + \mu_{\sigma} - \Sigma_{\sigma n} - E]
\nonumber \\[10pt]
& + &  \beta^{-1} \sum_{\sigma n} \ln G_{\sigma n}^{-1} - \beta^{-1}
\big\langle
\ln {\cal Z} \{ {\cal G}^{-1} , \epsilon_i \} \big\rangle_{av} \quad ,
\label{Gl6}
\end{eqnarray}
where $N(E)$ is the density of states of non-interacting electrons,
$\omega_n = (2n+1) \pi \beta^{-1} $ are Matsubara frequencies and
the generalized atomic partition function ${\cal Z}$
is represented by a Grassmann functional integral
\begin{mathletters}
\label{Gl7}
\begin{equation}
{\cal Z} \{ {\cal G}^{-1} , \epsilon_i\} =
\int {\cal D} \psi {\cal D} \psi^* e^{\cal A}
\label{Gl7a}
\end{equation}
with the local action
\begin{equation}
{\cal A} = \Bigg[ \sum\limits_{\sigma , n}  \psi_{\sigma n}^*
({\cal G}_{\sigma n}^{-1} - \epsilon_i )
\psi_{\sigma n}
 - U \int\limits_0^{\beta} d \tau
\psi^*_{\uparrow} (\tau) \psi_{\uparrow} (\tau)
\psi_{\downarrow}^* (\tau) \psi_{\downarrow} (\tau) \Bigg]
\label{Gl7b}
\end{equation}
\end{mathletters}
Here $\Psi_\sigma (\tau)$ are Grassmann
variables that depend on the imaginary time $\tau$, and
$\Psi_{\sigma n}$ are their Fourier
transforms into  $\omega_n$-space.
Only the random potentials $\epsilon_i$ are configuration
dependent. Furthermore,
\begin{equation}
{\cal G}^{-1}_{\sigma n} = G_{\sigma n}^{-1} + \Sigma_{\sigma n}
\label{Gl8}
\end{equation}
is an effective local propagator of the electrons.
The grand potential (\ref{Gl6}) is a functional of the complex quantities
$G_{\sigma n}$ and $\Sigma_{\sigma n}$ which at this point enter as
variational parameters. That is,
the {\em physical} values of $G_{\sigma n}$ and $\Sigma_{\sigma n}$
(namely, the local
Green function, $G_{\sigma n} \equiv G_{ii, \sigma n}$,
 and the self-energy, $\Sigma_{\sigma n} \equiv \Sigma_{ii, \sigma n}$,
 of the electrons, respectively) correspond
to those for which (\ref{Gl5}) is stationary
\cite{22,47}.

Each term of $\Omega_{av}$ in (\ref{Gl6}) has a clear
mean-field interpretation,
which allows us to outline a simple construction of $\Omega_{av}$ on a
mean-field level \cite{39,18}.
We
first introduce a homogeneous \cite{48},
 effective, energy-dependent potential $\Sigma_{\sigma n}$
which is defined in
such a way that in the
thermodynamic limit the system of disordered, interacting electrons
is equivalently described by a system of non-interacting electrons in
the potential $\Sigma_{\sigma n} $.
 The free energy
$\Omega_{av} \{ \Sigma_{\sigma n} \}$
 is now constructed
as follows \cite{49}: (i) we start with the free energy
of non-interacting electrons in the potential $\Sigma_{\sigma n}$
 (first  term in
 (\ref{Gl6})),
(ii) then we remove the potential $\Sigma_{\sigma n}$
 from site $i$, i.~e. subtract
its energy contribution (second term in (\ref{Gl6})),
and (iii) replace it by the  potential
$\hat{v}_{i \sigma} = \frac{1}{2} U \hat{n}_{i - \sigma}
+ \epsilon_i - \mu_\sigma$,
 and finally average over $\epsilon_i $ (last term in (\ref{Gl6}) )
\cite{50}. To determine the potential $\Sigma_{\sigma n}$
we demand that the free energy obtained
by this construction be stationary under variation of $\Sigma_{\sigma n}$,
 i.~e.
$\partial \Omega_{av}/\partial \Sigma_{\sigma n} = 0$.
This is a self-consistent equation
for $\Sigma_{\sigma n}$.
 The major advantage of the
mean-field grand potential (\ref{Gl6}) is the property
that the variational parameters are not explicitly configuration
 dependent \cite{39}, i.e.~they are determined
only by {\em averaged} quantities
which result  from the averaging in (\ref{Gl6}). Since the averaging
is local, the disorder only leads to local correlations on the same site.
Different lattice sites are effectively decoupled and all
the information about the surrounding sites is contained in the
effective potential  $\Sigma_{\sigma n}$.
Indeed,  $\Sigma_{\sigma n}$ is a generalization of the ``coherent
potential'' known from the theory of random alloys
\cite{40,41}. This may be seen as
follows \cite{18,51}:
if we neglect the Hubbard interaction $U$, the Grassmann
functional integral (\ref{Gl7a}) becomes Gaussian and
can be performed exactly, leading to
\begin{eqnarray}
 \Omega_{av} (U = 0)/L & = & - \beta^{-1} \sum_{\sigma ,n}
\big\{ \int_{- \infty}^{\infty}
 d E N(E) \ln [i \omega_n + \mu_{\sigma} - \Sigma_{\sigma n}
- E] \nonumber \\[10pt]
&+ & \langle \ln (1 + G_{\sigma n } (\Sigma_{\sigma n} - \epsilon_i) )
\rangle_{av} \big\}
\label{Gl9}
\end{eqnarray}
This is precisely the grand potential of a random alloy with non-interacting
electrons within CPA, with  $\Sigma_{\sigma n}$
as the coherent potential expressed in Matsubara frequencies
\cite{41,51}.
Hence (\ref{Gl6}) can be understood as a
field-theoretic generalization
of the CPA to random alloys with {\em interacting}
electrons which becomes exact for $Z \to \infty$.

 It should be noted
that the averaged grand potential (\ref{Gl6}) does  not contain the complete
information about the equilibrium physics of random
systems \cite{39}.
 For example, averaged {\em products} of Green functions contain new,
non-trivial correlations and cannot be derived from (\ref{Gl6}). The averaged
grand potential does, however, carry the full information about the
equilibrium {\em thermodynamics} of the model. We concentrate
in our paper exclusively on the thermodynamics
of the Anderson-Hubbard model.

It is well-known that, for large enough $U$,
 the Hubbard model itself leads to AFLRO
 at (or even close to)
half-filling.
This type of long range order must therefore be considered
even in the presence of disorder. For the sake of
simplicity we investigate only AFLRO
on bipartite lattices. In this case the breaking of symmetry
is caused by a staggered magnetic field $h_s$ with ($+h_s)$ on $A$- and
($-h_s$) on $B$-sites.
To be able to study this type of order
quantitatively we extend the averaged grand potential (\ref{Gl6})
to the symmetry-broken phase. The averaged
grand potential of a general antiferromagnetic solution has the form
\begin{eqnarray}
\hspace*{-20mm} \Omega_{av}/L  & = & - \frac{\beta^{-1}}{2 } \Big\{
\sum\limits_{\sigma, n}  \int_{- \infty}^{\infty} dE N(E) \ln
\left[ \;  ( i \omega_n + \mu_{A \sigma}- \Sigma_{A \sigma n} )
(i \omega_n + \mu_{B \sigma} - \Sigma_{B \sigma n} ) - E^2 \;
\right] \nonumber\\[10pt]
& - & \sum\limits_{\sigma , n}
\sum\limits_{\gamma = A,B} \ln G_{ \gamma \sigma n}^{-1}
+  \sum\limits_{\gamma= A,B} \;
\langle \ln {\cal Z}_{\gamma} \rangle_{av} \Big\}
\label{Gl10}
\end{eqnarray}
where the index $\gamma = (A,B) \equiv (+ 1, -1)$ corresponds to sublattices
$A, B$  respectively,
and $\mu_{\gamma \sigma} = \mu + \sigma(\gamma h_s +
 h)$, with $h$ as a  magnetic field.
The local partition function ${\cal Z}_{\gamma}$ in (\ref{Gl10}) is given by
(\ref{Gl7}) with the replacements
$\psi_{\sigma n} \to \psi_{\gamma \sigma n}, \psi_{\sigma} (\tau)
\to \psi_{\gamma \sigma}(\tau)$,
 $G_{\sigma n}^{-1} \to  G_{\gamma \sigma n}^{-1}$ etc.,
and integration over all Grassmann fields, i.e.~$ {\cal Z}_{\gamma} =
\int {\cal D} \psi {\cal D} \psi^* e^{{\cal A}_{\gamma}}$ with
\begin{equation}
{\cal A}_{\gamma} = \sum_{\sigma n} \psi_{\gamma\sigma n}^*
(G_{\gamma \sigma n}^{-1} + \Sigma_{\gamma \sigma n}
- \epsilon_i) \psi_{\gamma \sigma n} - U \int_0^\beta
d \tau \psi_{\gamma \uparrow}^* (\tau) \psi_{\gamma \uparrow} (\tau)
\psi_{\gamma \downarrow}^* (\tau) \psi_{\gamma \downarrow} (\tau).
\label{Gl11}
\end{equation}
The full partition function is given by ${\cal Z} = {\cal Z}_A {\cal Z}_B$.
As
described above the physical values of
$G_{\gamma \sigma n}$ and $\Sigma_{\gamma \sigma n}$ in (\ref{Gl10})
are then found from the stationarity conditions
\begin{equation}
\delta \Omega_{av}/\delta G_{\gamma \sigma n} = 0 \; , \;
\delta \Omega_{av}/\delta \Sigma_{\gamma \sigma n} =0 .
\label{Gl12}
\end{equation}
In the following we will set $h = 0$ in (\ref{Gl10}) and exclude the
existence of {\em ferri}magnetism.
In this case the average number of
electrons on sublattices $A$ and $B$ are equal.
Then there appears a new symmetry:
quantities with indices $\gamma, \sigma$ and $- \gamma, - \sigma$
coincide. Hence the double index
$\gamma, \sigma$ can be combined into a single one, $\alpha = (+, -),$
with $\alpha = +$ for $A\uparrow = B \downarrow$ and $\alpha
= - $ for $A\downarrow = B\uparrow$. Thereby the number of independent
parameters is reduced.
Eq. (\ref{Gl12}) then yields two coupled sets of
 self-consistent equations for
$G_{\alpha n }$ and $\Sigma_{\alpha n}$
\begin{mathletters}
\label{Gl13}
\begin{eqnarray}
G_{\alpha n } & = & \int\limits_{- \infty}^{\infty} dE
\frac{N(E)}{i \omega_n + \mu_\alpha - \Sigma_{\alpha n}
 - E^2/(i \omega_n +
\mu_{- \alpha} - \Sigma_{- \alpha n} ) } \label{Gl13a} \\[10pt]
G_{\alpha n} & = & - \int\limits_0^\beta d\tau \; e^{i \omega_n \tau}
\langle \langle \psi_{\alpha}
 (\tau) \psi_{\alpha}^* (\tau^+) \rangle_T \rangle_{av}
\label{Gl13b}
\end{eqnarray}
\end{mathletters}
where $\mu_\alpha = \mu + \alpha h_s$.
Here $\langle \ldots \rangle_T$ represents the
statistical average with the local
action
as
\begin{equation}
\langle {\cal O} \rangle_T  = \frac{1}{{\cal Z}_+ {\cal Z}_- }
\int {\cal D} \psi  {\cal D} \psi^* {\cal O} e^{{\cal A}_+ + {\cal A}_-}
\label{Gl14}
\end{equation}
Note that (\ref{Gl13a}) merely
expresses the fact that $G_{\alpha n}$
is the local element of the full Green function,
i.e.~is an explicit function of the self
 energy $\Sigma_{\alpha n}$. Only
eq. (\ref{Gl13b}) describes the actual
dynamics of the system,  determining
the self-energy $\Sigma_{\alpha n}$ as a function of the input parameters $T,
\mu, U$ and the disorder distribution.
Eqs. (\ref{Gl13}) fully determine the one-particle properties of the
Anderson-Hubbard model in $d = \infty$.

\section{Correlation functions and thermodynamic stability}

To decide about the stability of a particular solution of (\ref{Gl13}) we
have to evaluate averaged {\em two}-particle correlation functions or
susceptibilities.
Two such quantities are of particular interest in the Anderson-Hubbard
model: the averaged staggered magnetic susceptibility
$\chi_{av}^{stag} \equiv \chi_{AF}^{}$
and the averaged compressibility $\kappa_{av}$.
The former decides about the instability of the
paramagnetic phase w.r.t. AFLRO,
while the latter not only contains the information
about a possible instability w.r.t. phase separation ($\kappa_{av} \to
\infty$),
but also provides a thermodynamic criterion for a solution to be insulating
($\kappa_{av} \to 0)$. Quite generally susceptibilities can be obtained
from the second derivative of the averaged grand potential
$\Omega_{av}$ w.r.t. to some
variable $x$ as
\[
X_{av} = - \frac{1}{L} \; \frac{\partial^2\Omega_{av}}
{\partial x^2} =
\left\{
\begin{array}{ll}
\chi_{AF}^{} , & x = h_s 
\\[10pt]
\kappa_{av} ,&  x = \mu . 
\end{array}
\right.
\]
The first derivative of $\Omega_{av}$ w.r.t. $x$ yields
\begin{equation}
\frac{1}{L} \; \frac{\partial\Omega_{av}}
{\partial x} = - \frac{\beta^{-1}}{2} \sum_{\alpha n} f_\alpha^x
G_{\alpha n}
\label{Gl16}
\end{equation}
where $f_\alpha^\mu = 1$ and $f_\alpha^{h_s} = \alpha$. Differentiating
once more we obtain the desired susceptibilities
(15). Note that these susceptibilities
are evaluated at zero staggered field $h_s$, i.e.~$h_s$ is put to zero
once the second derivative of $\Omega_{av}$ w.r.t. $\mu$ or $h_s$ is taken
(in this case
$\mu_{\alpha} \equiv \mu$).
Thus the susceptibilities are given by
\begin{equation}
X_{av} = \beta^{-2} \sum_{\alpha n, \alpha' n'} f_{\alpha}^x
\Gamma_{nn', n'n}^{\alpha \alpha'} \gamma_{\alpha' n'}^x
\label{Gl17}
\end{equation}
where $\Gamma$ is the local two-particle correlation function
\[
\Gamma_{n_1n'_1 , n'_2 n_2}^{\alpha \alpha'} = \int_0^\beta
d \tau_1 d\tau'_1 d \tau_2 d\tau'_2 \exp
\big\{ i (\omega_{n_1} \tau_1 + \omega_{n'_1} \tau'_1
- \omega_{n'_2} \tau'_2 - \omega_{n_2} \tau_2) \big\}
\]
\begin{equation}
\times \big[ \langle \langle \psi_{\alpha} (\tau_1) \psi_\alpha^* (\tau_2)
\psi_{\alpha '}(\tau_1^\prime ) \psi_{\alpha'}^*
(\tau'_2) \rangle_T \rangle_{av}
- \langle \langle \psi_\alpha (\tau_1) \psi_\alpha^* (\tau_2) \rangle_T
\langle \psi_{\alpha '} (\tau'_{1}) \psi_{\alpha '}^* (\tau'_2 ) \rangle_T
\rangle_{av}  \big]
\label{Gl18}
\end{equation}
The quantity $\gamma_{\alpha n}^x = \partial (G_{\alpha n}^{-1} +
\Sigma_{\alpha n})/\partial x$ in (\ref{Gl17}) measures the response of
the averaged medium to an infinitesimal change of the
field $x$. We will later see that $\gamma_{\alpha n}^x$ decides about the
(in-)\linebreak[3]stability of a given phase.
This dynamical response function
is determined by an integral equation in frequency-space which does not
explicitly depend on momentum. (Note that there are no convolutions in
$\vec{k}$-space
in the $d = \infty$ limit as is typical for a mean-field
theory). This property does
not imply, however, that the response
function $\gamma_{\alpha n}^x$ is local, too. It
only indicates that $\gamma_{\alpha n}^x$ is {\em diagonal}
in the momentum $\vec{q}$. A momentum dependence will enter
by taking derivatives, e.g. of the local Green function, w.r.t.
an external field with a particular $\vec{q}$-dependence ($\vec{q} = 0$
in the case of the compressibility or the ferromagnetic susceptibility,
and $\vec{q} = (\pi , \ldots , \pi)$ for the staggered susceptibility).
 Infrared divergencies
show up in the spectrum of $\gamma_{\alpha}^x(t)$.

The integral equations for $\gamma_{\alpha n}^x$ are
derived as follows. Using
the effective propagator ${\cal G}_{\alpha n}^{-1} = G_{\alpha n}^{-1} +
\Sigma_{\alpha n } = (1+ G_{\alpha n} \Sigma_{\alpha n})/G_{\alpha n}$, such
that
\begin{equation}
\gamma_{\alpha n}^x = \partial {\cal G}_{\alpha n}^{-1}/\partial x
\label{Gl19}
\end{equation}
we replace the self-energy in (13) by $\Sigma_{\alpha n} =
{\cal G}_{\alpha n}^{-1}
- G_{\alpha \sigma n}^{-1}$. Then (13) reads
\begin{mathletters}
\label{Gl20}
\begin{eqnarray}
G_{\alpha n} & = & \langle g_{\alpha n} (E) \rangle_E \\
\label{Gl20a}
G_{\alpha n} & = & -\int_0^\beta d \tau e^{i \omega_n \tau}
\langle \langle \psi_{\alpha} (\tau) \psi_{\alpha }^* (\tau^+) \rangle_T
\rangle_{av}
\label{Gl20b}
\end{eqnarray}
\end{mathletters}
where $g_{\alpha n}(E) = (z_{\alpha n} - E^2/z_{-\alpha n})^{-1}$,
with $z_{\alpha n} = i \omega_n + \mu + \alpha h_s
- {\cal G}_{\alpha n}^{-1} + G_{\alpha n}^{-1} $,
and
\begin{equation}
\langle y (E) \rangle_E = \int_{\- \infty}^\infty dE \; N(E) \; y(E).
\label{Gl21}
\end{equation}
To obtain $\gamma_{\alpha n}^x$ we differentiate (\ref{Gl20}) w.r.t. $x$.
This yields
\begin{mathletters}
\label{Gl22}
\begin{eqnarray}
G^\prime_{\alpha n} & = &\left( \gamma_{\alpha n}^n +
\frac{G^\prime_{\alpha n}}
{G_{\alpha n}^2} - f_\alpha^x
 \right) \xi_{\alpha n} + \left( \gamma_{-\alpha n}^x
+ \frac{G^\prime_{-\alpha n}}{G^2_{-\alpha n}} - f_{- \alpha}^x \right)
\eta_{\alpha n}
\label{Gl22a} \\
G^\prime_{\alpha n} & = & \beta^{-1} \sum_{\alpha' n'}
\Gamma_{n n', n'n}^{\alpha \alpha '} \gamma_{\alpha ' n'}^x
\label{Gl22b}
\end{eqnarray}
\end{mathletters}
where $G'_{\alpha n} = \partial G_{\alpha n}/\partial x$ and
$\xi_{\alpha n} = \langle g_{\alpha n}^2 (E) \rangle_E, \eta_{\alpha n} =
\langle E^2 g_{\alpha n}^2 (E) \rangle_E/z_{- \alpha n}^2$. Eliminating
$G'_{\alpha n}$ from (\ref{Gl22}) we obtain the following matrix equation
 for $\gamma_{+,n}^x $
 and $\gamma_{- , n}^x$:
\begin{equation}
\beta^{-1} \sum_{\alpha' n'} (\beta \delta_{nn'} R_{n'}^{\alpha \alpha'} +
\Gamma_{n n', n'n}^{\alpha \alpha'} ) \gamma_{\alpha' n'}^x = v_{\alpha n}^x
\label{Gl23}
\end{equation}
The quantities $R_n^{\alpha \alpha'}$ and $v_{\alpha n}^x$ are a matrix and
a vector, respectively, w.r.t. the sublattice-spin
index $\alpha$ and may be written as
\begin{mathletters}
\begin{eqnarray}
\label{Gl24}
{\rm\bf R}_n & = & - (det {\rm\bf D}_n)^{-1} {\rm\bf D}_n {\rm\bf T}_n
\label{Gl24a}\\
\vec{v}_n^x & = &  {\rm\bf R}_n \cdot \vec{f}^x
\label{Gl24b}
\end{eqnarray}
\end{mathletters}
where
\begin{mathletters}
\label{Gl25}
\begin{equation}
{\rm\bf D}_n = \left(
\begin{array}{cc}
1- \frac{\xi_{-,n}}{G^2_{-,n}} & \frac{\eta_{+,n}}{G^2_{-,n}} \\[8pt]
\frac{\eta_{-,n}}{G^2_{+,n}} & 1- \frac{\xi_{+,n}}{G^2_{+,n}}
\end{array}
\right)
\label{Gl25c}
\end{equation}
\begin{equation}
{\rm\bf T}_n  =  \left(
\begin{array}{ll}
\xi_{+,n} & \eta_{+,n} \\[8pt]
\eta_{-,n} & \xi_{-,n}
\end{array}
\right)
\label{Gl25a}
\end{equation}
\begin{equation}
\vec{f}^x = \left(
\begin{array}{l}
1 \\[8pt]
f_-^x
\end{array} \right).
\label{Gl25b}
\end{equation}
\end{mathletters}
In the paramagnetic phase, where $f_\alpha^x v_{\alpha n}^x = f_{-\alpha}^x
v_{- \alpha n}^x$, eqs. (23)-(25) reduce to (\ref{Gl6}) in Ref.\cite{42}.

To determine the boundary between the paramagnetic and the antiferromagnetic
phase it is sufficient to calculate $\chi_{AF}$ only in the paramagnetic
phase.
By contrast, $\kappa_{av} $ has to be evaluated in {\em both}
phases since in both phases a metal-insulator transition (MIT) is
possible, in principle. Indeed, in the paramagnetic phase of a random alloy
with discrete disorder spectrum a MIT due to band-splitting
may occur (at $U = 0$ this is {\em known}
to be the case, at least within CPA
\cite{40,41}), and even in the presence of
AFLRO there are indications that both a metallic and an
insulating phase exist \cite{42}. We will find that the most sensitive
indicator for an incipient transition are the dynamic response functions
$\gamma_{\alpha n}^x, x = \mu, h_s$  -- especially their behavior close
to the Fermi energy (i.e.~their values for the lowest Matsubara frequencies),
rather than
$\chi_{AF}^{} $ and $\kappa_{av}$ itself. The latter
quantities are much less sensitive to changes in the ground state
since they represent {\em sums} over the Matsubara frequencies.

\section{Specification of the Model and of the Numerical Procedure}

Eqs. (15)- (\ref{Gl25}) form the basis
for the numerical evaluation of $\kappa_{av}$ and $\chi_{AF}^{}$.
Before they can be solved quantitatively we have to specify the
model parameters we use.
The numerical calculations were performed with a semi-elliptic
density of states (DOS)
with total width $2w$, i.e.
\begin{equation}
N(E) = \frac{2}{\pi w^2} (w^2 - E^2)^{1/2} .
\label{Gl26}
\end{equation}
This DOS is chosen because of its sharp algebraic band edges,
resembling those typical for $d = 3$, and its
simple analytic form; it is exact for a Bethe lattice in the limit
$Z \to \infty$.

To study the influence of the disorder we investigate, and compare,
three qualitatively different distributions of random potentials:\\
a)
\underline{Discrete, binary-random-alloy distribution}
\begin{mathletters}
\label{Gl27}
\begin{equation}
P_{binary} (\epsilon_i) = \frac{1}{2} \delta \Big( \epsilon_i -
\frac{\Delta}{2}\Big)
+ \frac{1}{2} \delta \Big( \epsilon_i + \frac{\Delta}{2} \Big) .
\label{Gl27a}
\end{equation}
The atomic potentials
$\epsilon_i = \pm \Delta/2$ are chosen with equal probability
to ensure an average band filling of $n = 1$ (only in this case, or
$n \simeq 1$, is AFLRO expected to occur at all). This
distribution is important since
it leads to a disorder-induced MIT due to
band-splitting in the
noninteracting system (an exact result in $d = \infty$ [38,39] )
which may compete with the interaction-induced Mott-Hubbard MIT.\\
\noindent
b) \underline{Continuous, semi-elliptic distribution}
\begin{equation}
P_{semi} (\epsilon_i) = \frac{8}{\pi \Delta^2} \Big[ \Big(
\frac{\Delta}{2} \Big)^2 - \epsilon_i^2 \Big]^{1/2}
\label{Gl27b}
\end{equation}
This is a much softer type of disorder.
We found that a continuous, {\em constant} disorder distribution
$P_{const} (\epsilon_i) = \Delta^{-1} \theta
(\frac{\Delta}{2} - | \epsilon_i |)$,
leads to essentially identical results.\\
\noindent
c)  \underline{Percolation-type disorder}
\begin{equation}
P_{perc} (\epsilon_i) = (1-x) \delta (\epsilon_i)
\label{Gl27c}
\end{equation}
\end{mathletters}
By using $P_{perc} (\epsilon_i)$ one may simulate doping in a non-random
system with impurities which do not hybridize with conduction electrons:
with probability $x$ an infinite energy
barrier is created which prevents the electrons from visiting
these sites.
The three distributions
 allow us to test
the universality of the magnetic behavior of the model obtained for
different types of disorder.

 To be able to study the competition between
magnetic order caused by the electronic interactions and the disordering
effects caused by the random potential, respectively,
 we work with an average band filling $n_{av} = 1$.
Due to the symmetry of the distributions (\ref{Gl27a},b)
we can fix the
chemical potential at $\mu = U/2$. This can even be done with (\ref{Gl27c})
 in which
case the Green function has to be weighted with an additional factor of $1-x$.

For the numerical evaluation of the functional integral (\ref{Gl14}) we
employ
the algorithm of Hirsch and Fye \cite{28}.  We discretize the time
variable, i.~e. $\beta = \Lambda \delta \tau$, with $0.25 \leq \delta
\tau \leq 1$, and then extrapolate the quantities under investigation to
$\delta \tau \to 0$.
To obtain a smooth imaginary-time Green function $G(\tau)$ even for
discrete values of $\tau$ we use instead of $G_n$ (the Fourier transform
of $G(\tau)$ w.r.t. Matsubara frequencies $\omega_n$) the function
\begin{mathletters}
\label{Gl28}
\begin{eqnarray}
\overline{G}_n & = & \delta \tau/ [1- \exp (\delta \tau /G_n)]
\label{Gl28a} \\
G(\tau) &= &\beta^{-1} \sum_n e^{i \omega_n \tau} \overline{G}_n
\label{Gl 28b}
\end{eqnarray}
\end{mathletters}
with $\beta = \Lambda \delta \tau$. In the continuum limit $\delta \tau
\to 0$ the functions $\overline{G}_n$ and $G_n$ coincide. The definition
(\ref{Gl28}) preserves the relation
$G(\tau = 0) = n - 1$ even in the case
of discrete $\delta \tau$ and suppresses
unphysical and undesirable oscillations in $G(\tau)$ due to
the discretization (see Fig.~1). In particular, by smoothing
$G(\tau)$ in this simple
way large-scale Fourier transformations (namely for every iteration)
described in Ref.~\cite{29} become unnecessary.

Exact summations over spin variables in the discrete Hubbard-Stratonovich
transformation were used whenever possible, i.~e. for $\Lambda \leq 22$.
For $\Lambda > 22$ we used Monte-Carlo sampling.
 After typically 4-8 iterations an accuracy of
$10^{-5}$ and $10^{-3}$ was reached in the exact summations and in the
Monte-Carlo sampling with $10^4$ sweeps per iteration, respectively.
Close to the magnetic transition the convergence becomes significantly slower
and the Monte-Carlo sampling less and less efficient.
This implies that in the
immediate vicinity of the transition this method cannot be used to
obtain accurate results.

The integration over the continuous disorder
 distributions was performed using the
Gauss-Legendre quadrature \cite{53}.
A discretization of the random energies in steps of $\delta \epsilon/w \sim
0.2$ gave an accuracy of $\sim 10^{-5}$ for all observables.
The required CPU time increases proportional to $\Lambda^2 2^\Lambda$ and
$\Lambda^3$ for the exact summation and the Monte-Carlo
sampling, respectively
\cite{52}. Although the random energies break
 the particle-hole symmetry we
never encountered a minus-sign problem.

Setting $\hbar = k_B =  1$ the only remaining physical dimension
is that of an energy ($U, \Sigma, \Delta, T,$ etc.) or
inverse energy ($G, \kappa_{av}, \chi_{AF}^{}, \beta, \delta \tau$).
Departing
from our earlier convention we now choose the {\em half-band width}
$w$ as our energy unit since it does not depend on the limiting process
$Z \to \infty$. (This is in contrast to the scaling of the hopping
amplitude $t = \alpha t^*/\sqrt{Z}$, where $\alpha$ may be chosen at will).
This convention agrees with that used by Kotliar and collaborators
\cite{21,23,30,32}.
To be able to compare the results presented in this paper with our
earlier ones \cite{42}, all numerical values of
quantities with dimension of energy
(inverse energy) obtained in Ref.~\cite{42} must be divided (multiplied)
by a factor of 2. To compare with the results  of Jarrell's group
\cite{24,25,33}
 and that of Georges and Krauth \cite{29,31},
 their numbers have to be divided
(multiplied) by a factor of $\sqrt{2}$.

\section{Magnetic Phase Diagram}

The competition between disorder and correlations in the Anderson-Hubbard
model strongly affects its properties both in the ground state and at
finite temperatures. Concerning the thermodynamics the influence of
different kinds of disorder on the stability of AFLRO near half filling
is of particular interest. To determine the instability of the paramagnetic
phase w.r.t. the formation of AFLRO we
evaluate eqs.~(15) - (25) in the paramagnetic limit for $x = h_s$, in which
case the dependence on the index $\alpha$
drops out; we set $\Sigma_{\alpha n} \equiv \Sigma_n$, etc. and
$\gamma_{\alpha n}^{h_s} \equiv \gamma_n$. Note that the matrix
equation (\ref{Gl23}) separates into two identical
scalar equations in this case. As already mentioned below (\ref{Gl18})
it is not the averaged susceptibility itself which is of primary interest
in the investigation of the stability of the paramagnetic phase but
the dynamic response function $\gamma_n$. The averaged
susceptibility diverges (becomes negative) if and only if the real
part of the response function $\gamma_n$ diverges for at least
one frequency. This is characteristic for the dynamics of a quantum
system of interacting particles possessing infinitely many coupled
internal degrees of freedom labelled by the Matsubara frequencies.
Each frequency corresponds to one mode in the quantum mechanical system which
is described by complex variables.
 Note, however, that the Matsubara frequencies do not
index the actual independent modes, since $\Gamma_{n n', n'n}$ is
not diagonal in frequency.
Only the eigenvalues of the integral equation (\ref{Gl23}) represent
the independent (eigen-)modes of the interacting quantum system.
We can ascribe a critical (N\'{e}el) temperature $T_{N, |n|}$ to each
eigenmode $n$ whereby $T_{N,|n|} > T_{N,|n'|}  $ if $|n| < |n'|$.
In Fig.~2 we show the function $\gamma_n (\tau)$, being a
linear combination of eigenmodes for the lowest Matsubara frequencies
which lie very close to the diverging independent soft mode. The
functions $\gamma_n (\tau)$ diverge due to the contribution of the eigenmode
with
the highest $T_{N,|n|} $. The highest critical temperature
$T_{N,|n |} $ is the thermodynamic critical temperature $T_N$.

Once the function $\gamma_n$ and the local vertex function $\Gamma_{n n',
n'n}$
are known one may calculate the averaged susceptibility.
Above $T_N$ this susceptiblity must obey the Curie-Weiss law, provided
there is such a $T_N > 0$. In Fig.~3 three characteristically different
 temperature dependencies of $\chi_{AF}^{-1}$ are shown
for one value of (binary alloy) disorder, $\Delta =2$, at different
interactions strengths. If $U/\Delta$ is sufficiently large the
Curie-Weiss law
with a finite $T_N$ is obeyed as in the case without disorder.
For values of $U \sim \Delta$ the low-temperature behavior begins to be
determined by the scattering off the frozen random configurations and
deviations from the Curie-Weiss-law become apparent. For $U/\Delta \ll1$ we
observe a minimum in $\chi_{AF}^{-1}$ (i.e.~a maximum in $\chi_{AF}$
itself) which separates the temperature-dominated regime from the
disorder-dominated regime. In the latter case the long-range
correlations are continually suppressed due to impurity scattering.
The phase diagram in the $T-U$-plane calculated from the
zeros of $\chi_{AF}^{-1}$ is plotted in Fig.~4a for different
values of the binary alloy disorder. We can distinguish two different
regimes,
a) \underline{$U < U_c \simeq 2.5:$} the disorder gradually suppressed the
long-range order and thus reduces the critical temperature. This is exactly
what one expects from scattering off frozen random configurations; b)
\underline{$U > U_c$}: here the situation is strikingly  different
since $T_N$-curves of constant disorder start to {\em cross}.
This means that a small amount of disorder {\em supports} the
formation of AFLRO, i.e.~the critical temperature {\em increases}
with disorder as shown in Fig.~5. A maximal
critical temperature is reached at some value $\Delta_c (U)$
beyond which a further increase of disorder causes the critical temperature
to decrease monotonically to zero.
This effect is particularly pronounced at strong coupling.
It is also observed, but less pronounced,
in the case of the continuous disorder distribution (Fig.~4b).
In the limit $U \gg \Delta, t$ the enhancement of $T_N$
may be explained as follows \cite{54}.
The virtual hopping of an electron with spin $\sigma$
from a given site $A$ with local energy $\epsilon_A$ to a
neighboring site $B$ with energy $\epsilon_B$ occupied by a ($-\sigma$)
electron leads to an energy gain $J_1 = -t^2/[U-(\epsilon_A - \epsilon_B)]$
and to $J_2 = -t^2/[U + (\epsilon_A - \epsilon_B)]$ for
the reverse process. The effective spin coupling $J(\epsilon_A , \epsilon_B)$
is given by the sum of these energies, $J (\epsilon_A , \epsilon_B) =
J_1 + J_2$, which for a bounded disorder distribution, $- \frac{\Delta}{2}
\leq \epsilon_{A,B} \leq \frac{\Delta}{2}$, with $U \gg \Delta$
(strong coupling) implies
\begin{equation}
J(\epsilon_A , \epsilon_B) = \frac{2t^2}{U} \left[1 + \left(
\frac{\epsilon_A - \epsilon_B}{U} \right)^2 \right]
\label{Gl29}
\end{equation}
Assuming that even in the presence of disorder the N\'{e}el-temperature
$T_N (U, \Delta) \linebreak[4]
\mbox{$\propto < J (\epsilon_A, \epsilon_B) >_{av}$}$,
 we find
\begin{mathletters}
\begin{equation}
\frac{T_N (U, \Delta)}{T_N (U,0)} = \int d\epsilon_A \int d \epsilon_B
\; J (\epsilon_A, \epsilon_B) P(\epsilon_A) P(\epsilon_B)
= 1 + \frac{2}{U^2} [p_2 - p_1^2 ]
\label{Gl29a}
\end{equation}
for arbitrary disorder distribution $P(\epsilon)$. $p_1$ and $p_2$ are the
first and second moment of $P(\epsilon)$, where $p_l = \int d \epsilon
\epsilon^l P(\epsilon)$. For a symmetric, bounded distribution $(p_1 =0)$
one therefore finds
\begin{equation}
\frac{T_N (U, \Delta)}{T_N (U,0)} =
1 + \lambda \left( \frac{\Delta}{U} \right)^2
\label{Gl29b}
\end{equation}
\end{mathletters}
with $\lambda = 2p_2/\Delta^2$, i.e.~the disorder is indeed found to
{\it increase} $T_N$ irrespective of the type of disorder. This effect is the
more
pronounced the more structure $P(\epsilon)$ has at $ | \epsilon |
\stackrel{<}{\sim} \frac{\Delta}{2}$, i.e.~is larger for
the binary alloy $(\lambda = \frac{1}{2})$ than for the semi-elliptic
$(\lambda = \frac{1}{8})$ distribution.
 This quadratic increase of $T_N$ with $\Delta$
is indeed found numerically at large $U$ (Fig.~5).
 Apparently the mechanism that
enhances $T_N$ is effective already at $U$-values as small as $U \sim 2.5$.
It is interesting to note that for given strength of disorder $\Delta$
the $U$-value where AFLRO begins to set in, $U_{min}$, given by $T_N
(U_{min}, \Delta ) = 0$, tracks almost
perfectly with $\Delta$, i.e.~$U_{min} (\Delta)
\simeq \Delta$. More generally, for binary alloy disorder
 the $T_N$-curves are found to obey the phenomenological scaling law
\begin{equation}
T_N(U, \Delta ) \simeq T_N \left( U \left[ 1- \left(
\frac{\Delta}{U} \right)^2 \right],0 \right)
\left[ 1- \frac{1}{2} \left( \frac{\Delta}{U} \right)^2 \right]
\label{Gl30}
\end{equation}
for {\it arbitrary} $U$ (Fig.~6). For $U \gg \Delta$ (31) reduces to (29b).

It is well-known that  AFLRO may be destroyed
by doping the (non-random) system with
holes, such that $n_{av} < 1$. This  is a very interesting effect both
from an
experimental and theoretical point of view. The effect of
adding holes may be approximated by introducing percolation-type
disorder, (\ref{Gl27c}), where the random site-energy is
zero with probability $1-x$ and infinite with probability $x$. Since
in $d = \infty$ the critical value for percolation is $x = 1$
\cite{55} AFLRO will persist up to $x = 1$. The boundaries
between the paramagnetic and the antiferromagnetic phases for this
type of disorder are shown in Fig.~7.
At weak coupling the critical temperature remains constant over a
wide range of $x$ and then rapidly falls to zero. At strong coupling,
however,
the critical temperature
{\it monotonically} decreases and approaches the linear behavior
$T_N (x) = T_N (0) (1-x)$. The latter dependence of $T_N$
which is very different from the one observed in the previous two cases
is an exact result
in the case of the disordered Falicov-Kimball model \cite{39}.

The magnetic correlations are mediated by the local magnetic moments. Their
static average, $m_{av}$, is defined by
$m_{av}^2 \equiv L^{-1} \sum_i \langle \langle
(\hat{n}_{i \uparrow} - \hat{n}_{i \downarrow})^2 \rangle_{T}\;
\rangle_{av} = 1 - 2 d_{av}$. Here $d_{av} = L^{-1} d \Omega_{av}/dU$ is
the average double occupancy of lattice sites.
For the binary and the semi-elliptic
 disorder distributions discussed above $m_{av}^2$ is shown
as a function of $U$ in Fig.~8a.
The disorder is seen to have two main effects
which are independent of the specific disorder distribution:
1) at a fixed value of $U$ an increase of the disorder {\em reduces}
the moments. \linebreak
2) For fixed disorder strength $\Delta$ an increase of $U$
leads to an increase of the moments,
with saturation starting at $U \stackrel{>}{\sim} \Delta$. Both features are
easily explained in terms of the effect the disorder and the on-site
repulsion, respectively, have on the average double occupancy $d_{av}$.
For $U \ll \Delta$ the local repulsion is weak while the spatial fluctuations
of the atomic potentials are strong, such that $d_{av}$ is at its maximum
value ($d_{av} \simeq 1/2)$ and $m_{av}$ is small. As $U$ increases
the particles are forced
to separate and $d_{av}$ decreases, i.e.~$m_{av}$ increases, too.
The local magnetic moments do not show a
critical behavior close to the transition
temperature $T_N$.
In the case of the percolation-type disorder (Fig.~8b) the density of
local moments on the reduced lattice with $(1-x)L$ sites increases
with concentration $x$. In other words, in the presence of disorder,
the saturation of $m_{av}^2$ sets in at smaller
values of $U$ than in the pure
system. The influence of the percolation-type disorder can
be explained by the reduction of the kinetic energy by a factor
of $1-x$ which is due to
the reduced average number of nearest neighbors. For large $U$,
in the Heisenberg limit, the factor $t^2$ in the antiferromagnetic
coupling, $J \sim t^2/U$, has to be replaced by
$t^2(1-x)$ leading to a linear decrease of $T_N$. For small values
of $U$, the influence of the repulsion increases due to the reduction
of the kinetic energy. Therefore, the double occupancy is
suppressed, i.e.~$m_{av}^2$ is enhanced by the percolation
disorder. The enhancement of $m_{av}^2$ leads to the stabilization
of AFLRO. This effect is definitely not described within the
Hartree-Fock approximation; the latter only leads to an
exponential suppression of $T_N$ if $U$ or $(1-x)$ becomes small.

In Fig.~9 the temperature dependence of the order parameter
(the averaged staggered magnetization $M_{AF}$) is shown. In the vicinity
of the transition point, $T \stackrel{<}{\sim} T_N$, it is well
represented by a mean-field-type dependence $M_{AF} (T) \propto
 (T_N - T)^{1/2}$ indicated
by the dashed line. We observe that the extrapolation of the Monte-Carlo
data to $M_{AF} (T_N)= 0$ using this law leads to the
{\it same} critical temperature
as that obtained from the divergence of
$\chi_{AF}^{}$, i.e.~$\chi_{AF}^{-1} (T_N)
= 0$, assuming a Curie-Weiss law for $T \stackrel{>}{\sim} T_N$.
Deviations from the square-root behavior of $M_{AF} (T)$ set in at lower
temperatures.

\section{Metal-Insulator Transitions}

In the preceding section we found that at weak coupling the disorder
suppresses the AFLRO. In the ground state the system is then a disordered
paramagnet. Whether it is metallic or not depends on the type and
strength of the disorder. In $d = \infty$, where Anderson localization
does not occur [38,39] a paramagnetic insulator only forms if the
spectrum of the disorder distribution is multiple connected, such as in the
case of the
binary-alloy disorder. For this type of randomness a MIT due to band
splitting is expected to occur at some value $\Delta \geq 1$. In the
non-interacting
case, $U = 0$, the exact result in $d = \infty$ (obtained by CPA \cite{40})
is $\Delta = 1$.

For $U =0$ there are two equivalent criteria to decide on whether
the ground state in $d = \infty$ is metallic or insulating. They correspond
to (i) a {\em spectral} definition of an insulator, based on
the disappearance of the DOS at the Fermi level (assuming that only
extended electrons are present), and (ii) a {\em thermodynamic} definition,
employing the disappearance of the compressibility \cite{56}. However, in
interacting systems these two criteria need not coincide. Hence
we will investigate both. In Fig.~10a the averaged compressibility
$\kappa_{av}$
of interacting electrons in the presence of binary-alloy disorder is shown
as a function of $U$ for $\beta = 16$. Hence $\kappa_{av}$ is calculated
across the
line $T = 0.0625$ in Fig.~4a, i.e.~within the paramagnetic and the
antiferromagnetic
phase, respectively (the transition points are indicated by arrows). For a
given
$\Delta \stackrel{>}{\sim} 1$ the curves display a common behavior
as a function of $U$: $\kappa_{av}$ is (exponentially) small at $U = 0$, then
increases
and, at $U \simeq \Delta$, approaches a maximum, beyond which  it becomes
(exponentially) small again for $U \gg \Delta$. This behavior has a clear
physical
interpretation: For $U \ll \Delta$, with $\Delta \stackrel{>}{\sim} 1$,
the double occupancy of lattice sites is at its maximum value
($d_{av} \simeq \frac{1}{2}$), at the same time the kinetic energy cannot
delocalize these
states. Hence the system is insulating.
At $T = 0$ and $U = 0$ this is an exact property for $\Delta > 1$ (split-band
limit) in $d = \infty$.
 As $U$ increases the on-site Coulomb repulsion forces the
particles to separate from each other. As a consequence the particles become
less localized. We then expect that, at some critical value
$U_c^{MI,1}$, a macroscopic fraction of the electrons becomes extended
and the system starts to be metallic. This real-space picture has its analogy
in $\vec{k}$-space (or w.r.t. energy): as the interaction increases the
previously
separated bands change their shapes, i.e.~the interaction-induced
energy exchange between particles leads to a transfer of states
into the energy gap where the Fermi energy is located. At the
critical value $U_c^{MI,1}$ the (still algebraic) band edges
reach the Fermi level, producing a finite DOS there as well as a finite
overall compressibility. (We cannot rule out that $U_c^{MI,1} = 0$, i.e.~that
the DOS at the Fermi level is finite even at arbitrarily small $U$; however,
this
would require the bands to acquire exponential tails -- a feature which
cannot be
observed numerically). Hence, in contrast to an interacting system without
disorder the Coulomb interaction in a disordered system is able to
{\em improve}
the metallicity of the system and may even turn an insulator into a
metal.
Apparently the interaction-induced energy exchange smooths  the energy
spectrum of
the disordered system, thus leading to an easier transfer of energy.

For $U > U_c^{MI,1}$ the system is then expected to be
a metal without Fermi liquid
properties (since $Im \Sigma (\omega\!=\!0) \neq 0$), at least in
$d = \infty$. As $U$
is further increased the DOS and $\kappa_{av}$ increase, reaching
a maximum at $U \simeq \Delta$. For even larger $U$
the effect of pushing states into the gap is reversed and
$\kappa_{av}$ decreases again. Hence at some critical value
$U_c^{MI,2}$ a {\em second} transition occurs, back into an insulating state,
where doubly occupied sites are almost completely suppressed, whereby the
mobility
is obstructed by the repulsive interaction.

In Fig.~10b $\kappa_{av}$ for the continuous, semi-elliptic disorder
distribution
is shown. Here $\kappa_{av}$ behaves qualitatively as in the non-random case,
i.e.~$\kappa_{av}$ decreases monotonically since the DOS does not split at
$U = 0, T = 0$. This
reduction of $\kappa_{av}$ occurs in two steps: for $U \stackrel{<}{\sim}
\Delta$ $\kappa_{av}$ decreases almost linearly with $U$, corresponding to
the linear
suppression of doubly occupied sites $d_{av}$, or $m^2_{av}$ (see Fig.~8a);
in this region
the system is metallic but not a Fermi liquid. Then, for $U \stackrel{>}
{\sim}\Delta$, when the number of doubly occupied sites is almost zero
so that the local moments are almost saturated, $\kappa_{av}$
approaches zero (exponentially) slowly. Strictly speaking, at finite
temperatures one has $\kappa_{av} > 0$ for $U < \infty$.

Next we discuss the relation of $U_c^{MI,1}$ and $U_c^{MI,2}$ to the
critical
interaction strength $U_c^{AF}$ where the paramagnet becomes unstable w.r.t.
AFLRO. In Fig.~10a,b the position of $U_c^{AF}$ is indicated by an arrow.
Since the local moments vanish in the (binary-alloy)-disorder-induced
insulating phase, the insulator-to-metal transition at $U_c^{MI,1}$ occurs
{\em before} the magnetic order sets in, i.e.~$U_c^{MI,1} < U_c^{AF}$.
This raises the question of what happens as one goes through the magnetic
transition into the ordered phase: will the compressibility jump
to zero discontinuously or will there be an {\em antiferromagnetic metal},
i.e.~does the interaction-induced MIT coincide with the magnetic transition
or not? To answer this question the behavior of $\kappa_{av}$ near the
transition
point has to be investigated. In Fig.11a the change of $\kappa_{av}$ due to
the onset of AFLRO in the case of binary-alloy disorder is shown. The results
for the paramagnetic phase are compared with those for the ordered phase.
 At $\Delta = 0$ the AFLRO is seen to
suppress $\kappa_{av}$ drastically. By contrast,
at finite disorder ($\Delta = 2)$, this difference
almost vanishes. Apparently the disorder {\em stabilizes} the metallic
state close to $U_c^{AF}$. A similar behavior is observed for the continuous
disorder (Fig.~11b). These results suggest that in the vicinity
of $U_c^{AF}$ the system is an {\em antiferromagnetic metal}. However,
MC-techniques are not able to decide whether this is true even at $T = 0$.
We attempted to extrapolate our finite-temperature data to $T = 0$ but
could not find a simple, accurate
extrapolation law. In Fig.~12 the averaged compressibility
for the system with binary-alloy disorder, calculated at $T = 1/16$
and $1/40$, is shown. Although at the lower temperature $\kappa_{av}$
is lower and the slope is slightly steeper at the transition point, the
critical point (indicated by an
arrow) seems to be well inside the metallic phase.
In particular, the tail behavior for $U > U_c^{AF}$ does not show any
significant change.

To investigate whether, and how,
 an insulator evolves  it is instructive to plot the dynamical
response function $\gamma_{\alpha n}$, or the quantities
\begin{mathletters}
\begin{equation}
\kappa_n:=
\frac{1}{2} \sum_\alpha \partial G_{\alpha n}/\partial \mu
\label{Gl32a}
\end{equation}
 with
\begin{equation}
\kappa_{av} = \beta^{-1} \sum_n \kappa_n ,
\label{Gl32b}
\end{equation}
\end{mathletters}
as a function of Matsubara frequency $\omega_n = \pi T(2n +1)$. In Fig.~13a
the
real part of $\kappa_n$ is shown for a system without disorder, $\Delta = 0$,
for $U = 1.75$ and $T = 1/64$ (the system is then well inside the
antiferromagnetic
region, see Fig.~4a). The behavior of  $Re \kappa_n$ close to $\omega_n = 0$
is seen to be very different for the antiferromagnetic and the
(hypothetical) paramagnetic
solution. For the paramagnetic solution, which is metallic,  $Re \kappa_n$
is a
monotonically decreasing function of $| \omega_n|$. By contrast,
in the antiferromagnetic (i.e.~insulating) phase this is only so for
$| \omega_n | \stackrel{>}{\sim} 1$, while for $\omega_n$ close to zero
 $Re \kappa_n$becomes strongly negative. In the former case
the sum over $\kappa_n$, i.e.~$\kappa_{av}$, is then clearly positive,
while in the latter case $\kappa_{av}$ becomes very small due to the negative
contributions close to $\omega_n = 0$. This behavior at $\Delta = 0$ is now
contrasted with that in the presence of binary-alloy disorder
($\Delta = 2,$ see Fig.~13b) at $U = 2.65$ for two different temperatures:
$T = 1/20$ (close to the magnetic transition) and $T = 1/64$ (well inside
the ordered phase; see Fig.~4a). At $T = 1/64$ the range of $\omega_n$values
for which
$Re \kappa_n < 0$ is now even narrower the in the case without disorder.
This shows that a definite answer to the question of whether the
antiferromagnetic
phase close to $U_c^{AF}$ is insulating $(\kappa_{av} = 0)$ or not at $T = 0$
can only be obtained from the behavior of
$Re\kappa_n$ at very small $\omega_n$, i.e.~from $\kappa_{av}$ at very
low temperatures $(T < 10^{-2})$. Within the MC-approach
used here such low temperatures  are not attainable.

The situation is similar in the case of $- Im G_n$, the imaginary part
of the one-particle Green function $G_n = \frac{1}{2}
\sum_\alpha G_{\alpha n}$,
which yields a {\it spectral} condition for an
insulator (for $n = 0$ and in the limit
$T \to 0$ the function $- Im G_0 \pi$ coincides with the DOS at the
Fermi level). If $Im G_0 \to 0$ for $T \to 0$
the ground state of the system is insulating. In Fig.~14 the influence of
the onset of AFLRO on $-Im G_0$ is shown. Although in the disordered
system the difference between the magnetic and non-magnetic solution is now
slightly
stronger than in the case of $\kappa_{av}$ (see Fig.~13a) the magnetic
transition occurs where $Im G_0$ is large, i.e.~well inside the metallic
phase. In Fig.~15 the behavior of $- Im G_n$ at the few lowest values
of $\omega_n \geq 0$ is plotted for the same values of $U$ and $\beta$
as for  $Re \kappa_n$ in Fig.~13. In the case without disorder
(Fig.~15a) this behavior is qualitatively similar to that of $Re \kappa_n$
(Fig.~13a): the paramagnetic solution is clearly metallic while the
antiferromagnetic
phase is insulating since $ Im G_0 \to 0$. In the presence
of disorder (Fig.~15b), however, the  downturn of $- ImG_n$ in the phase
with AFLRO
($\beta = 64)$
sets in only very close to $| \omega_n| \simeq 0$ i.e.
at very low temperatures. If close to the transition the
antiferromagnetic phase were really insulating the decrease of
$- Im G_n$ will have to be very rapid. In Fig.~16 we plotted our
Monte Carlo data for $- Im G_0$ in the case of binary-alloy
disorder with $\Delta = 2$ as a function of temperature for several values
of $U$. Both at very small $U$ (i.e.~$U < 1$, where the disorder-induced
gap of the DOS dominates the behavior) and at large $U$ (Heisenberg-limit)
the temperature dependence may be clearly extrapolated to
$-Im G_0 = 0$ for $T=0$, i.e.~the ground state is insulating. However,
for $U \simeq U_c^{AF} (\Delta, T = 0)$, with $U_c^{AF} (2,0)\simeq
2-2.5$, these results cannot be safely extrapolated. Hence the
question whether in the disordered system close to the magnetic transition
the antiferromagnetic phase is metallic or not
 remains open.
As in the case of $\kappa_{av}$ one would have to go to temperature
$T \stackrel{<}{\sim} 10^{-2}$ to be able to decide whether the ground state
 of
the disordered system with AFLRO is metallic or insulating.

\section{Conclusions}

In this paper we presented a detailed, quantitative study of the physical
effects caused by the simultaneous presence of interactions and
randomness in a system of lattice electrons. To this end we investigated
the Anderson-Hubbard model with diagonal disorder  at half
filling in the limit of infinite spatial dimensions, i.e.~within a
dynamical mean-field theory, for three different
disorder distributions. Numerical results
were obtained by employing Quantum Monte Carlo techniques that provide an
explicit finite-temperature  solution of the model in $d = \infty$. No
further approximation were used.

To construct the thermodynamic phase diagram we derived and evaluated the
appropriate averaged two-particle correlation function, i.e.~a dynamical
 response function, whose poles determine the magnetic instabilities of the
disordered, interacting system. Only this function -- and not the averaged
staggered susceptibility $\chi_{AF}^{}$ itself,
which is only a weighted sum of the
response function over the (Matsubara) frequencies -- fulfills a
closed equation that determines the two-particle spectrum.
The value of the response function at the lowest frequency
is the most sensitive indicator for an instability of the system.
In the temperature dependence of $\chi_{AF}^{}$ two distinct disorder
regimes are observed: (i) for weak disorder the Curie-Weiss law holds,
while (ii) at strong disorder $\chi_{AF}^{}$ acquires a maximum at
a temperature below which a crossover from the temperature- to the disorder-
dominated regime takes place.

We demonstrated that at low temperatures and sufficiently strong interaction
there always exists a phase with antiferromagnetic long-range order
(AFLRO). Furthermore we discovered a new strong-coupling anomaly,
namely  that the N\'{e}el-temperature $T_N$ is not always
a monotonously decreasing function of disorder. Indeed, at strong
coupling and not too large disorder $T_N$ is always found to be an
{\it increasing} function of disorder, i.e.~disorder favors
the formation of AFLRO in this regime. This implies the existence of an
unexpected {\it disorder-induced} transition to a phase with AFLRO. Under
the assumption that $T_N$ is proportional
to the effective exchange coupling between
spins even in the disordered system we proved that for diagonal disorder the
anomalous behavior is  generic, i.e.~is independent of the type
of disorder distribution. It is a consequence of
the fact that for diagonal disorder the difference between the local
energies of neighboring sites becomes {\it smaller}
on average, thus leading to stronger effective exchange coupling.

We then studied the existence of
metal-insulator transitions in the Anderson-Hubbard
model. Although in $d = \infty$ Anderson localization does not take place
the presence of disorder may well have  other strong effects. In particular,
binary-alloy disorder is able to cause band-splitting (thereby resembling
the effect of genuine interactions) and
hence may induce a metal-insulator
transition all by itself. Special attention was given to the question
 whether or not the disorder allows for the stabilization of an
antiferromagnetic (AF) {\it metal}.
To this end the average compressibility $\kappa_{av}$
was evaluated both in the paramagnetic (P) and AF phase.
Contrary to our expectation the presence of disorder was found to
enhance the metallicity of the AF-phase close to the
P-AF transition. This enhancement
strongly suggests the existence of an AF metal in
the low-temperature phase of the Anderson-Hubbard mode.
To investigate whether this
AF phase persists to be metallic down to $T=0$,
at least in $d = \infty$,  we studied the
frequency components of the one-particle Green function and of $\kappa_{av}$,
respectively, down to $T = 1/64$,
our lowest temperature. However, a reliable
answer to this question can only be found at
still lower temperatures, which at present are beyond the
reach of the finite-temperature Monte Carlo techniques used here.
In any case, the transition scenario involving metal-insulator
and P-AF transitions obtained for the Anderson-Hubbard model is remarkably
rich. For alloy-type disorder with $\Delta > 1$ (split-band limit for
$U=0$) at $T=0$ an increase of the interaction $U$ from zero will
probably first lead to a transition from a paramagnetic insulator to a
paramagnetic metal at $U_c^{MI,1}$, then,
at $U_c^{AF}$, to an antiferromagnetic metal
and finally, at $U_c^{MI,2}$, to an antiferromagnetic
insulator. No compelling evidence was found
that $U_c^{AF}$ and $U_c^{MI,2}$ coincide.
In the case of a continuous disorder distribution one has
$U_c^{MI,1} = 0$ since band splitting never occurs.

The above findings prove that the interplay
between electronic interactions and scattering
from disorder leads to interesting, and even novel, physical effects.
In particular, the strong coupling anomaly discovered here call for an
experimental verification.

\noindent
{\large\bf Acknowledgments:}

We are grateful to A. Altland, P. van Dongen, M. Jarrell,
 H. M\"{u}ller-Krumbhaar and especially
E. M\"{u}ller-Hartmann for very useful
discussions and comments. This work was supported in part by the
Sonderforschungsbereich 341 of the Deutsche Forschungsgemeinschaft.

\newpage
%
\vskip-1cm
\par
\noindent
{\large\bf Figure Captions }
\medskip
\par

\begin{description}

\item[Fig.1] Local Green function $G(\tau)$ for $U=0$ and
$\mu=0$ (a), $\mu=0.5$ (b) obtained by usual Fourier transformation
(dotted line) and by redefinition acc. to eq.(13) (solid line)

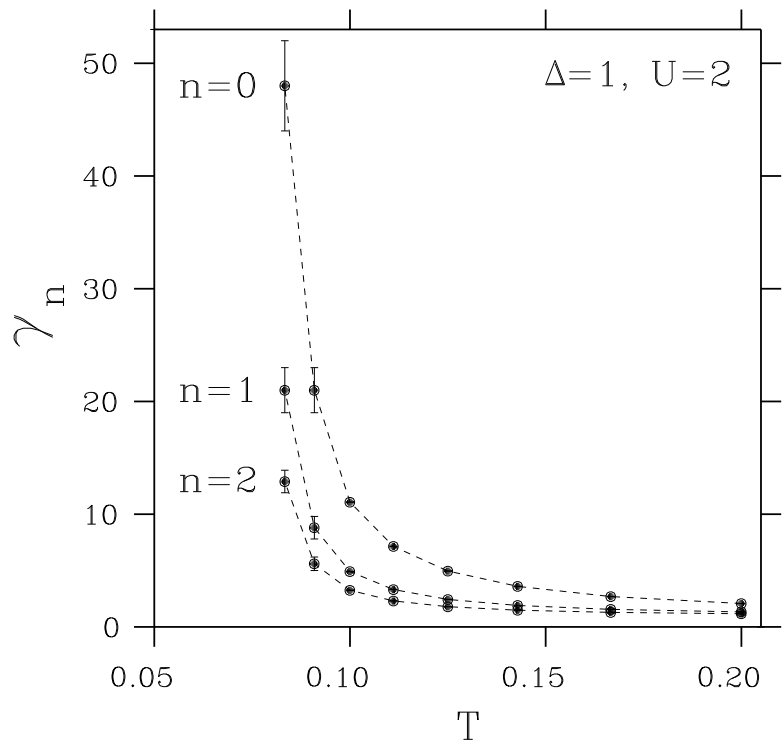
\item[Fig.2] Dynamical antiferromagnetic response function $\gamma_n$
vs.~temperature $T$. Binary alloy with disorder strength
$\Delta=1$ and $U=2$.
Here and in the following figures lines are usually guides for the
eye, and error-bars are roughly of the size of the symbols unless
shown explicitly.

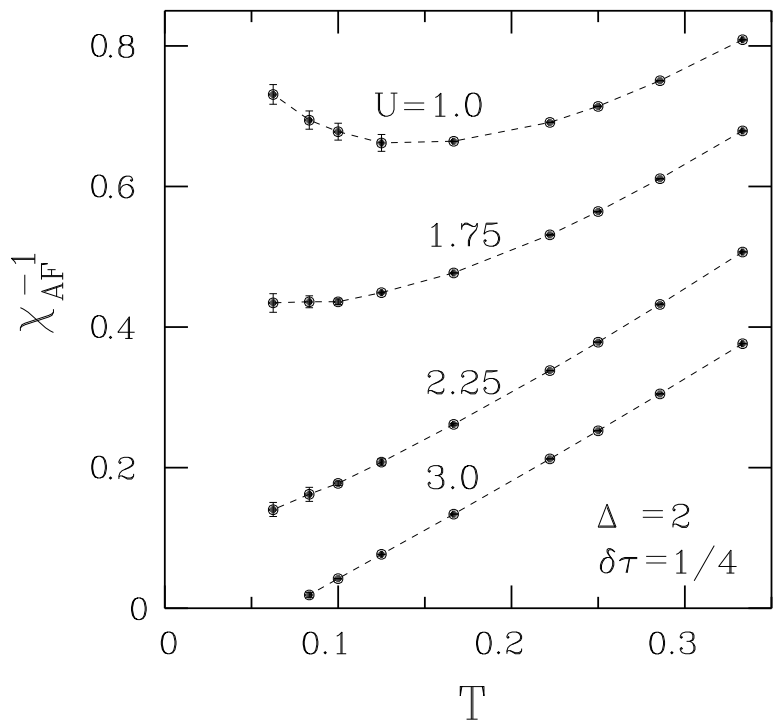
\item[Fig.3] Inverse averaged antiferromagnetic susceptibility
$\chi^{-1}_{AF}$ vs.~$T$ for the binary alloy with
$\Delta=2$ and several values of $U$.

\item[Fig.4a] $T-U$-phase diagram for the binary alloy with
$\Delta=0,1,2,4$ obtained from the zeroes of $\chi^{-1}_{AF}$
(see Fig.~4). The AF-phase is stable below the curves.
The dotted lines at $T=0$ depict the regimes where the Curie
law would give negative transition temperatures. Below the crosses
$\chi^{-1}_{AF}$ has no zeroes but a minimum and an AF-phase can
no longer be expected.

\item[Fig.4b] $T-U$-phase diagram for the semi-elliptic distribution
of the random energies with width $\Delta=0,2,4,6$.

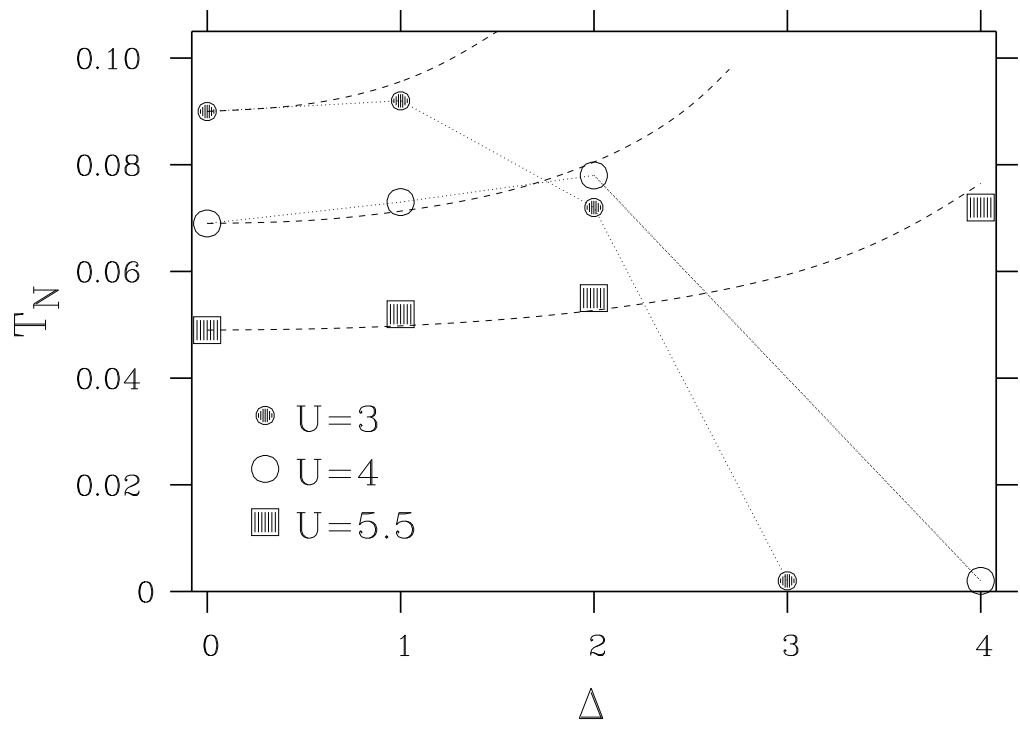
\item[Fig.5] $T$ vs.~$\Delta$ for the binary-alloy distribution at
$U=3,4$ and 5.5. Dashed lines: quadratic increase of $T_N$ according to
eq.~(30b) with $\lambda = 1/2$; dotted lines are guides to the eye only.

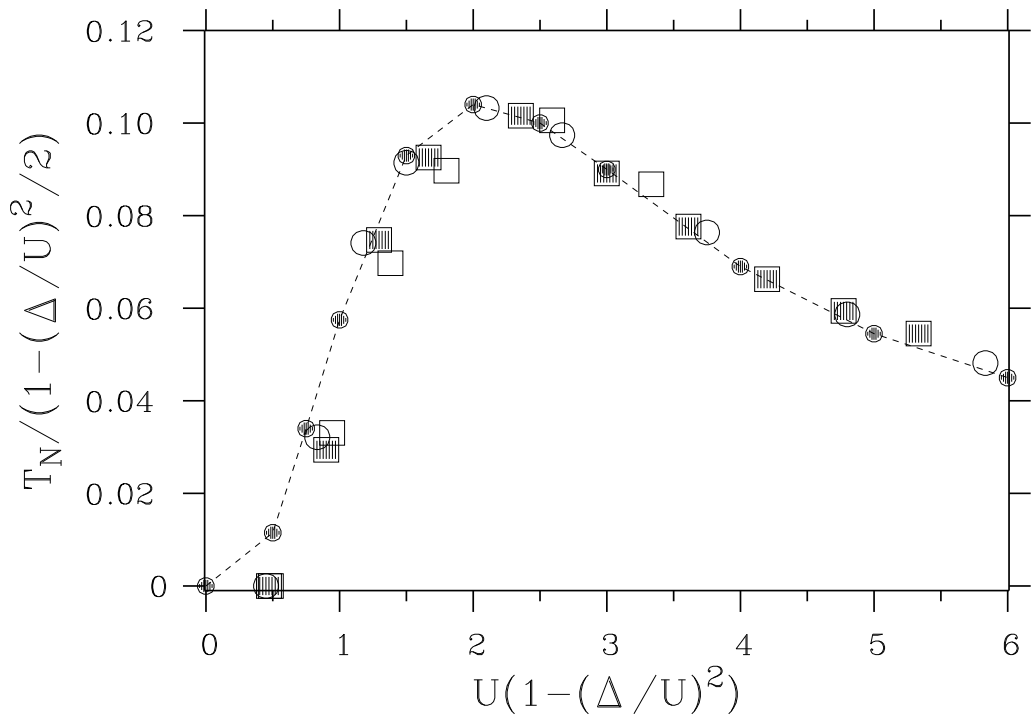
\item[Fig.6] Scaling plot of $T_N $ for the binary-alloy
 distribution according to eq.~(30); symbols as in Fig.~4a.

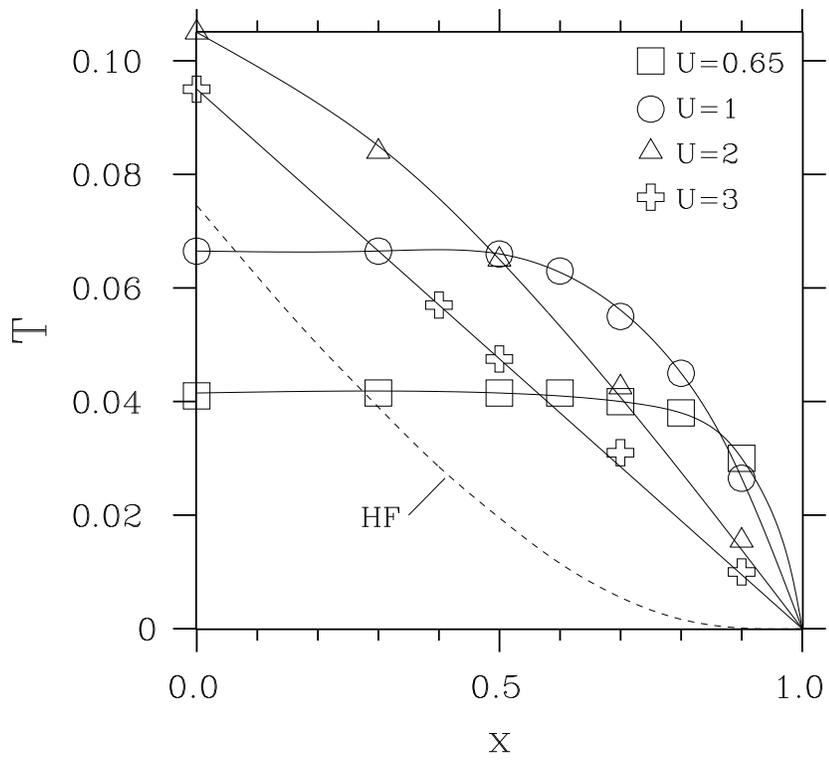
\item[Fig.7] $T-x$-phase diagram for the percolation-type disorder
for several values of $U$. The dashed line depicts the Hartree-%
Fock (HF) result for the smallest value of $U=0.65$.

\item[Fig.8a] Averaged quadratic local moment $m_{av}^2$ vs.~U
at inverse temperature $\beta=16$.
Without disorder (full circle); binary alloy with $\Delta=2$
(full square); semi-elliptic distribution with $\Delta=2$ (open circle)
and $\Delta=4$ (open square).

\item[Fig.8b] Averaged quadratic local moment $m_{av}^2/(1-x)$
(normalized to the concentration of sites with random energy zero)
vs.~$x$ for the percolation-type disorder at $\beta=16$ for several
values of $U$.

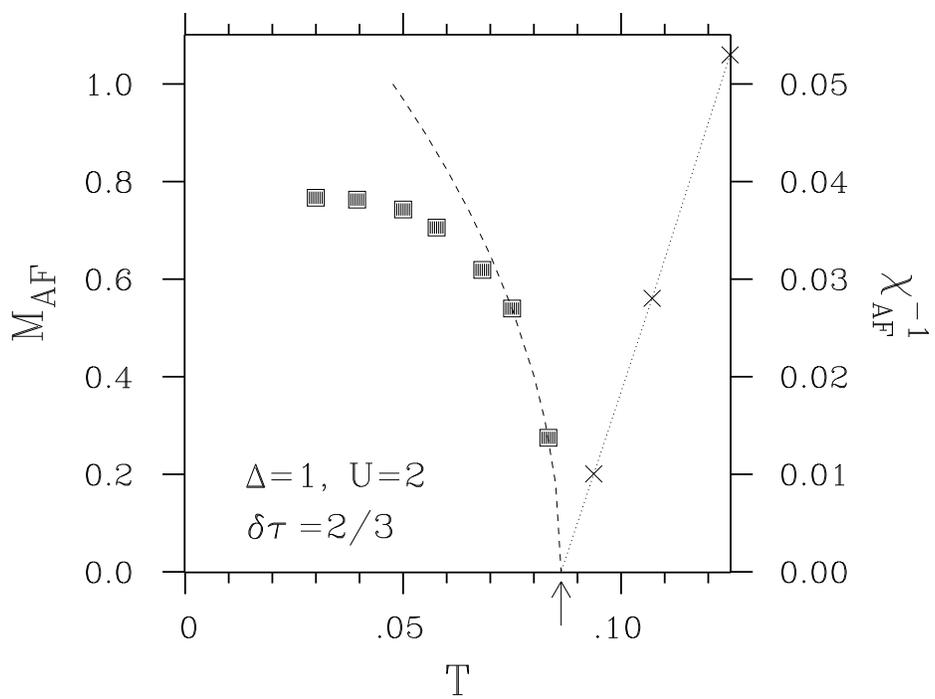
\item[Fig.9] Staggered magnetization $M_{AF}$ and
inverse averaged susceptibility $\chi^{-1}_{AF}$ vs.~$T$ for the
binary alloy with $\Delta=1$, $U=2$, $\delta\tau=2/3$.
Dashed line: square-root fit of the last two points of $M_{AF}$
below the transition; dotted line: linear fit of $\chi^{-1}_{AF}$.
The arrow indicates the extrapolated N\'eel temperature.

\item[Fig.10] Averaged compressibility $\kappa_{av}$ vs.~$U$ at
$\beta=16$. a) binary alloy, $\Delta=0,1,2,4$; b) semi-elliptic
distribution with $\Delta=0,2$ and 4.  Arrows indicate the
transition to the antiferromagnetic state.

\item[Fig.11] Averaged compressibility $\kappa_{av}$ vs.~$U$ at
$\beta=16$ in the paramagnetic phase (dashed lines)
and in the antiferromagnetic phase (dotted lines).
a) binary alloy, b) semi-elliptic distribution.

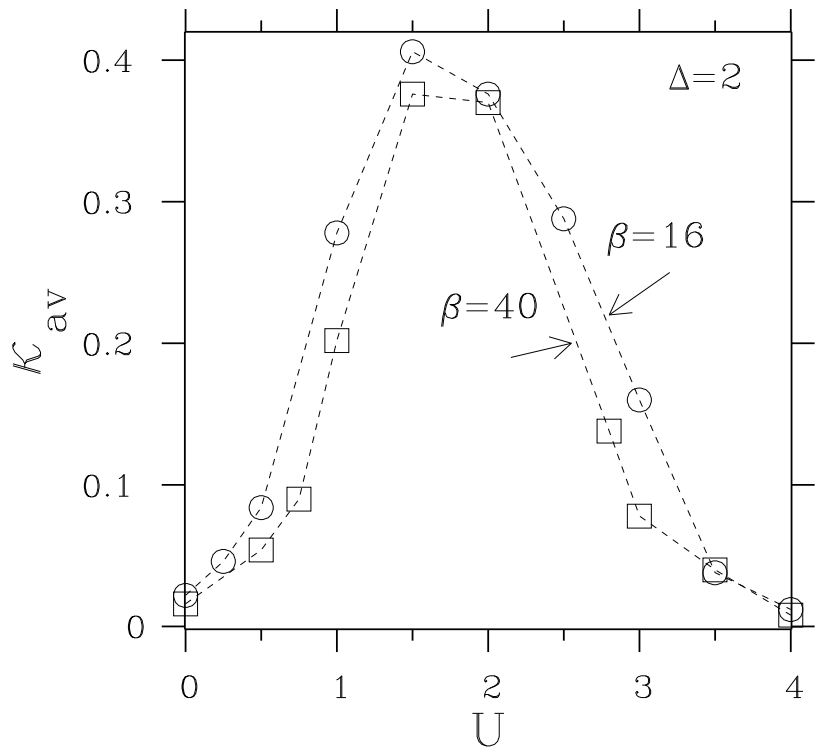
\item[Fig.12] Averaged compressibility $\kappa_{av}$ vs.~$U$
for the binary alloy at $\beta=16$ and 40. Arrows indicate the
transition to the antiferromagnetic state.

\item[Fig.13] Real part of $\kappa_{n}$ vs.~$\omega_n$ ,
a) for
$\Delta=0$, $U=1.75$ and $\beta=64$ in the paramagnetic phase (P)
 and the antiferromagnetic phase (AF); b)
 for
the binary alloy with $\Delta=2$, $U=2.65$
in the AF-phase at $\beta=40$ and 64.

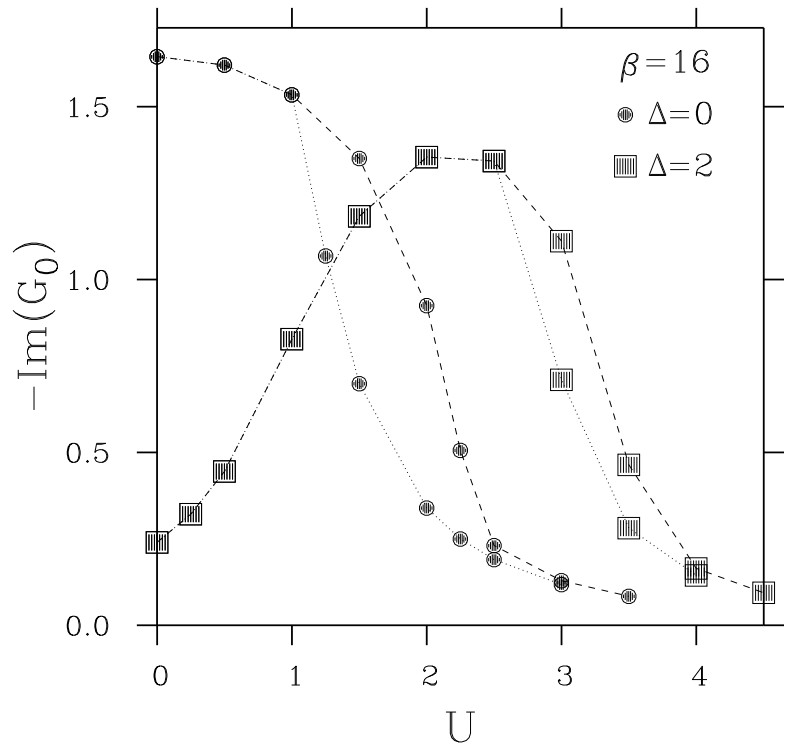
\item[Fig.14] Imaginary part of the one-particle Green function
$G_n$ at the lowest Matsubara frequency $\omega_0=\pi T $ vs.~$U$
for a system without disorder ($\Delta = 0$) and
with binary-alloy disorder ($\Delta = 2)$ at $\beta=16$;
P-Phase (dashed lines), AF-phase (dotted lines).

\item[Fig.15] Imaginary part of $G_n$ vs.~$\omega_n$ in the P-
and AF-phase; a) no disorder, U=1.75, $\beta=64$;
b) binary alloy with $\Delta=2$, $U=2.65$, $\beta=20,64$
(the two curves at $\beta=20$ in the middle
are undistinguishable).

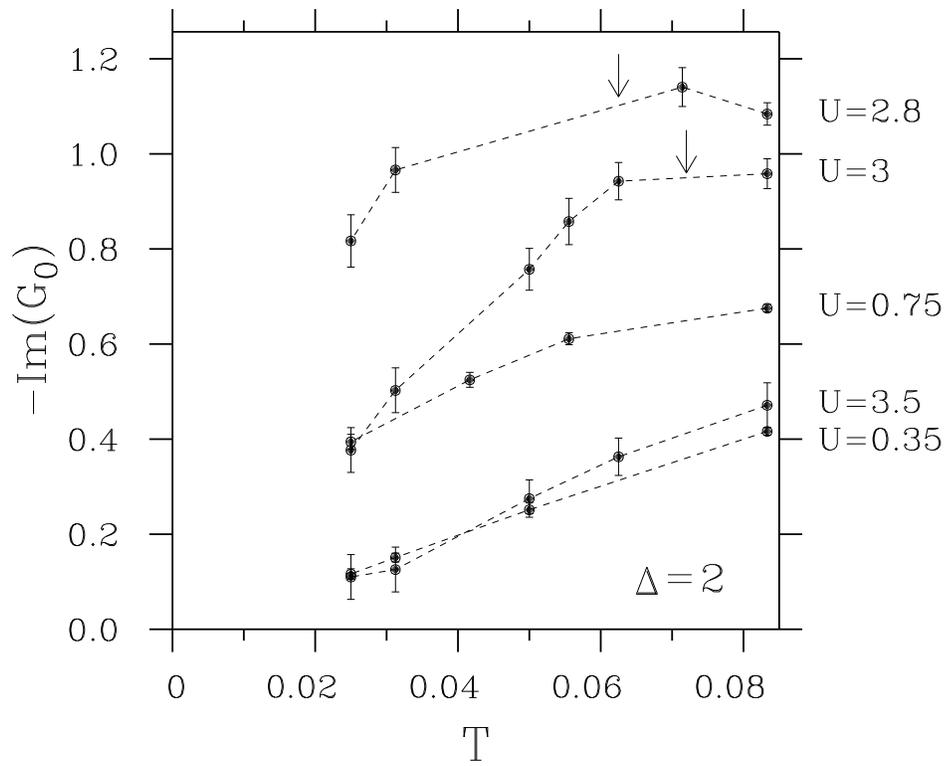
\item[Fig.16] Imaginary part of $G_0$ vs.~$T$ for the binary alloy
with $\Delta=2$ for several values of $U$.

\end{description}

\vfil\eject

\begin{thebibliography}{99}
\bibitem{49} Here we consider $\Omega_{av}$ as a functional of
$\Sigma_{\sigma n}$
only, i.e.~we assume that we already established the identity of
$G_{\sigma n}$ as the local Green function (see the discussion in ref.~47).
\bibitem{50} Note that the energy at site $i$ is measured {\em relative}
to the surrounding medium; that is why the effective local propagator of
electrons
at site $i$ is given by  ${\cal G}_{\sigma n} = [  G_{\sigma n}^{-1} +
\Sigma_{\sigma n}]^{-1}$.
\bibitem{52} All calculations were performed on a Cray YMP of the
Forschungszentrum J\"{u}lich.
\bibitem{54} This was pointed out to us for the binary-alloy case
by E. M\"{u}ller-Hartmann.
\end{thebibliography}
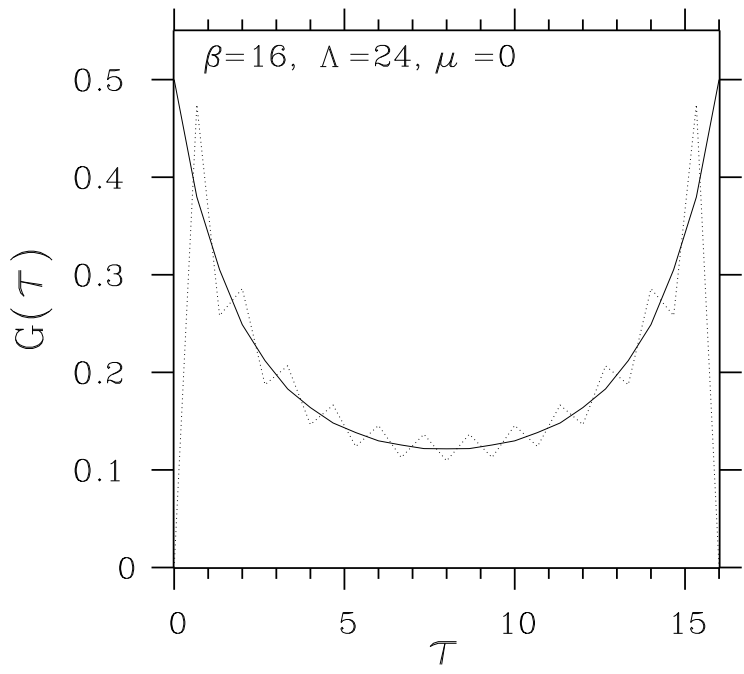
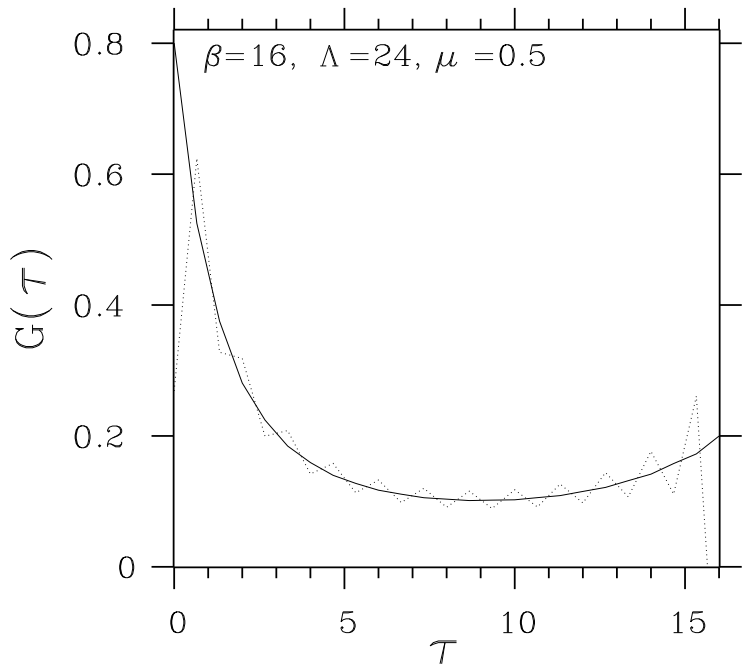
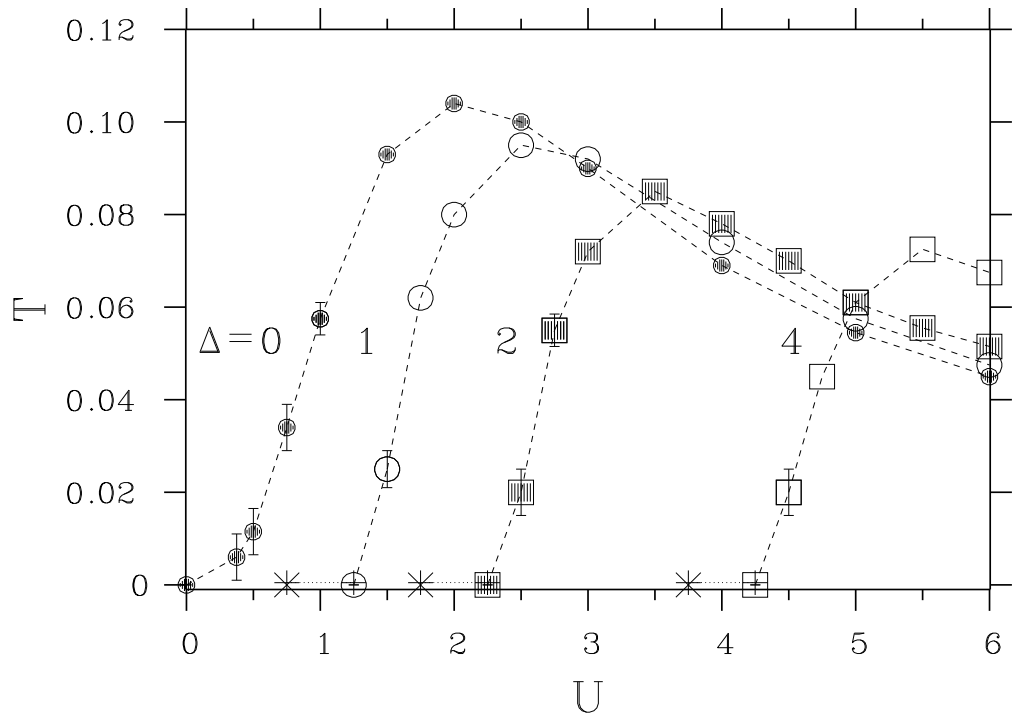
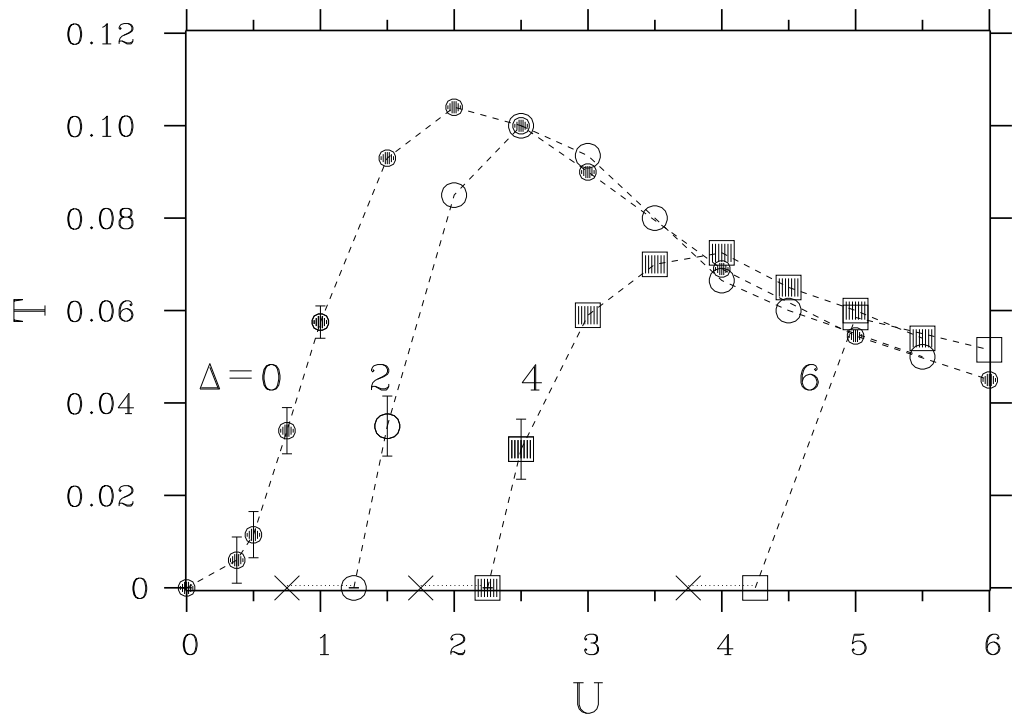
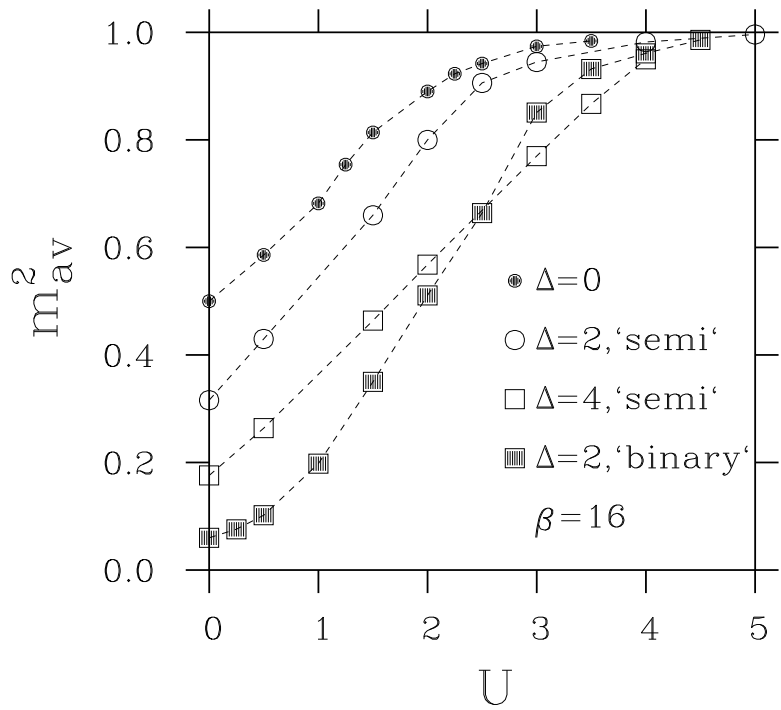
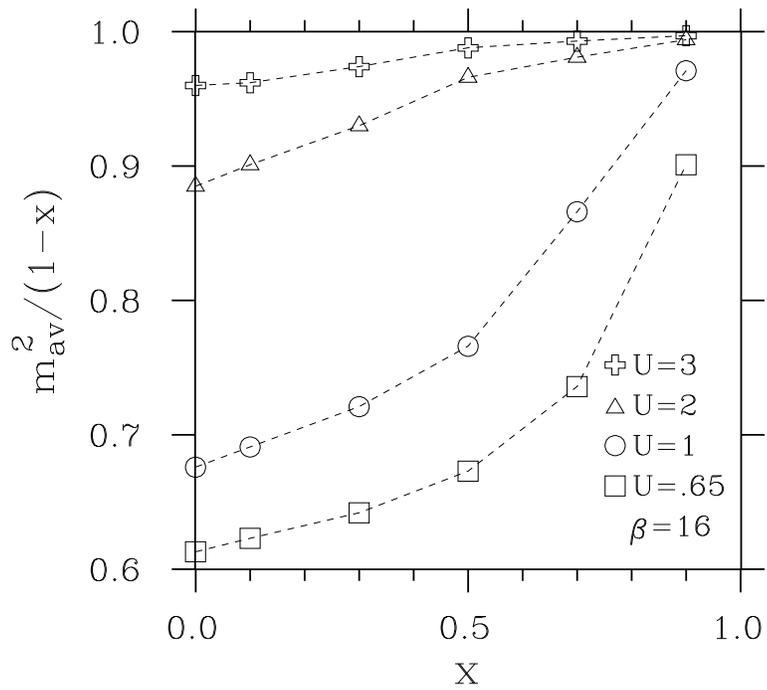
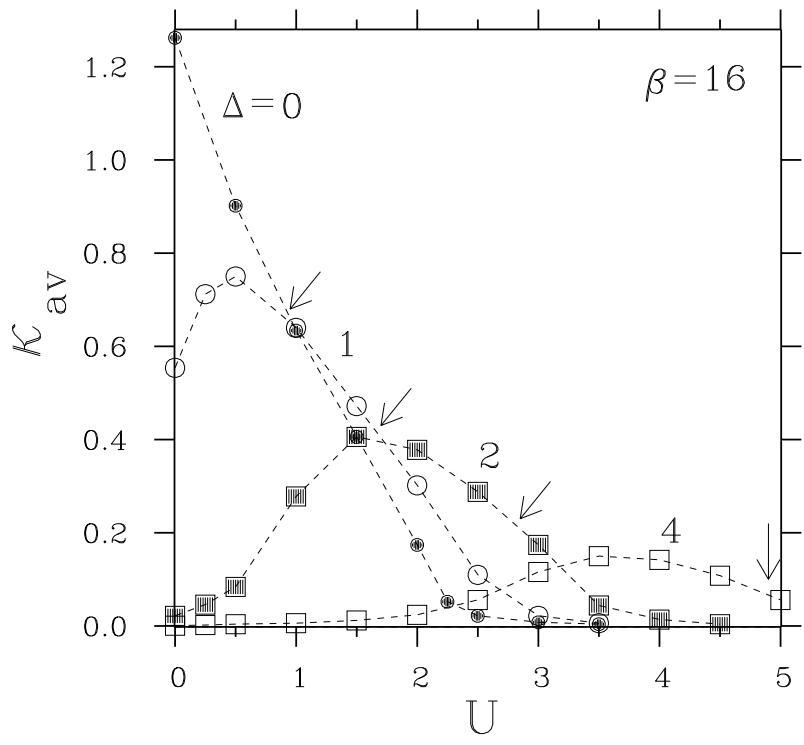
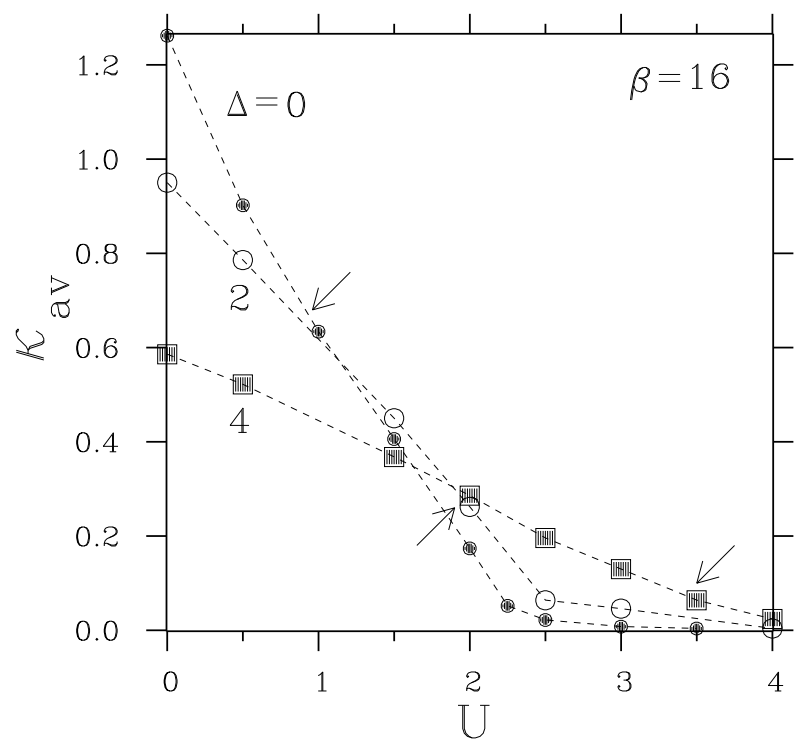
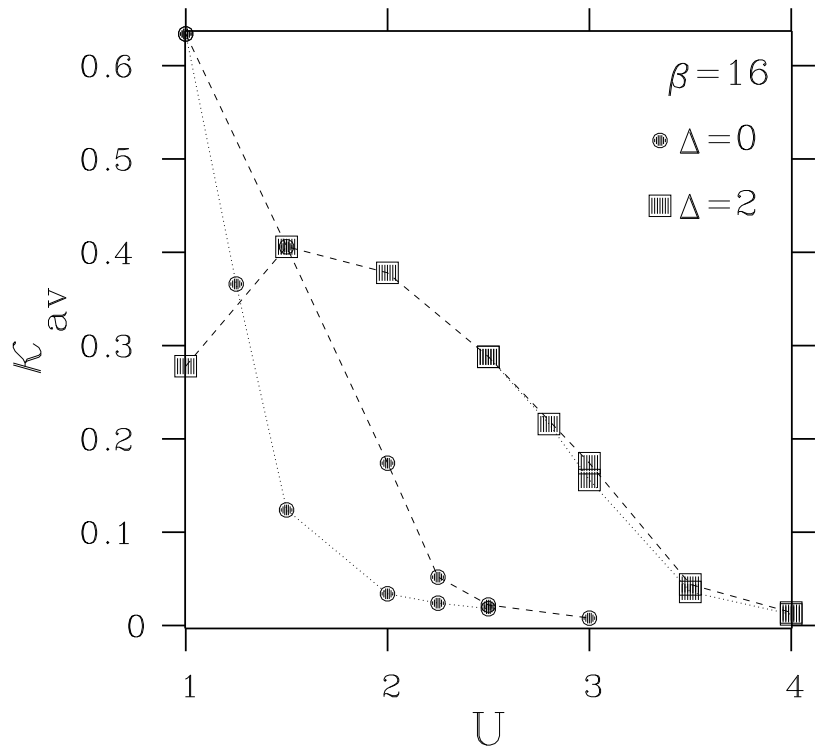
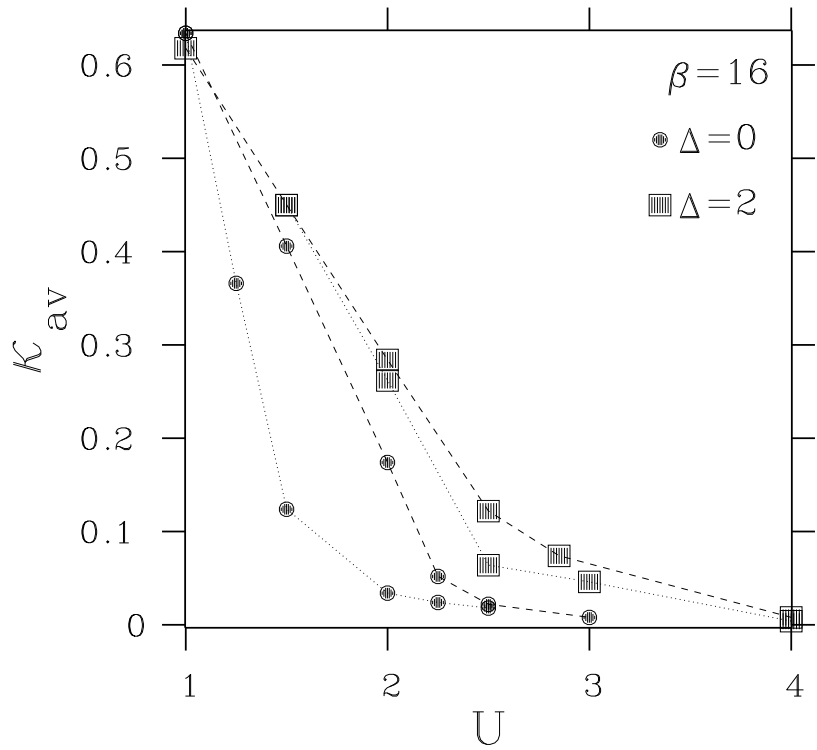
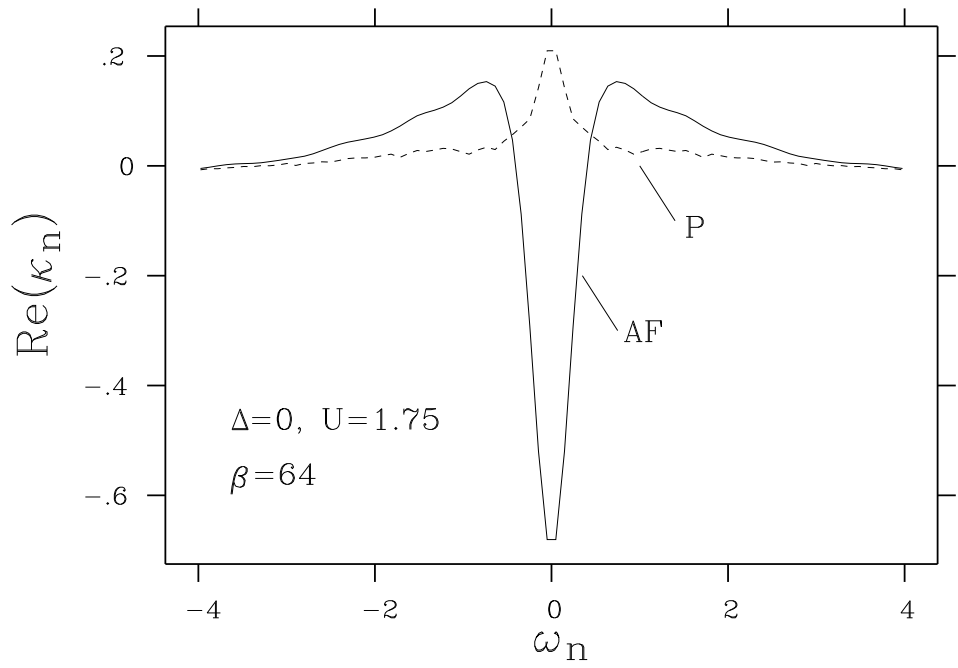
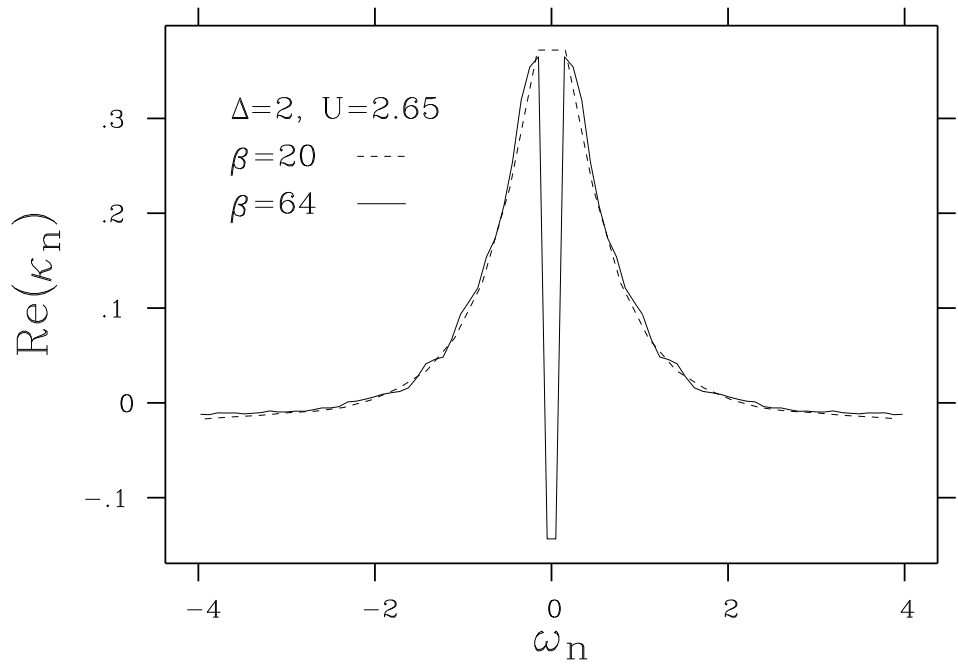
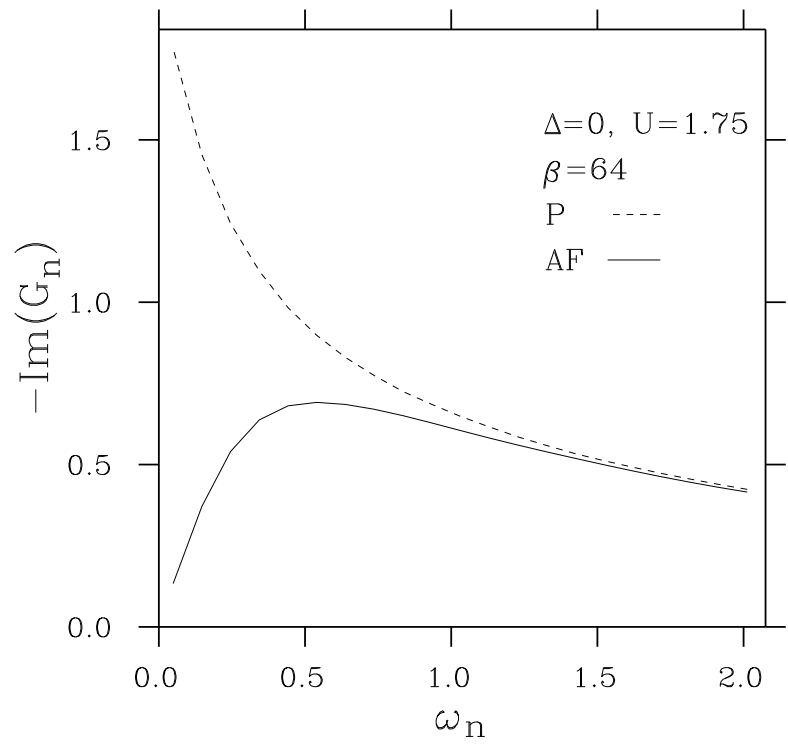
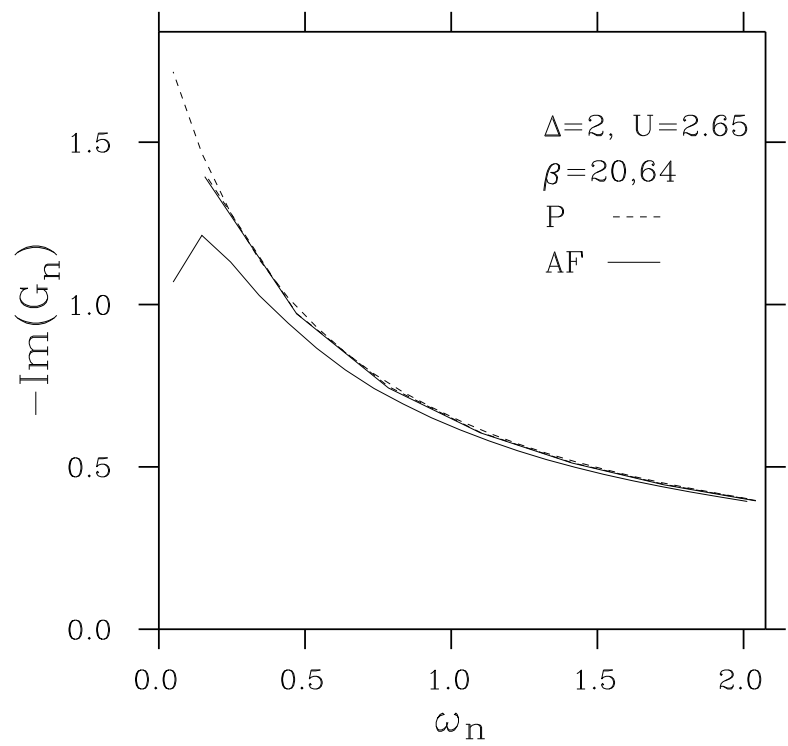
\end{document}